\documentclass[iop,usenatbib,useAMS,numberedappendix,tighten,appendixfloats]{emulateapj}
\bibpunct{(}{)}{;}{a}{}{,}
\usepackage{apjfonts}
\usepackage{enumerate,amsmath}

\newcommand{\ldust}{\hbox{$L_{\mathrm{dust}}$}}

\newcommand{\tcmbo}{\hbox{$T_\mathrm{CMB}^{\,z=0}$}}
\newcommand{\tcmb}{\hbox{$T_\mathrm{CMB}$}}
\newcommand{\tdusto}{\hbox{$T_\mathrm{dust}^{\,z=0}$}}
\newcommand{\tdust}{\hbox{$T_\mathrm{dust}$}}
\newcommand{\tkin}{\hbox{$T_\mathrm{kin}$}}
\newcommand{\tkino}{\hbox{$T_\mathrm{kin}^{\,z=0}$}}

\newcommand{\mic}{\hbox{$\mu$m}}
\newcommand{\lsun}{\hbox{$L_\odot$}}
\newcommand{\msun}{\hbox{$M_\odot$}}
\newcommand{\mdust}{\hbox{$M_\mathrm{dust}$}}

\newcommand{\nuco}{\hbox{$\nu_\mathrm{\textsc{co}}$}}

\newcommand{\tex}{\hbox{$T_\mathrm{exc}^{J_u}$}}

\def\lesssim{\mathrel{\hbox{\rlap{\hbox{\lower5pt\hbox{$\sim$}}}\hbox{$<$}}}}
\def\gtrsim{\mathrel{\hbox{\rlap{\hbox{\lower5pt\hbox{$\sim$}}}\hbox{$>$}}}}
 
 \DeclareMathSymbol{\la}{3}{AMSa}{46}
 \DeclareMathSymbol{\ga}{3}{AMSa}{38}

\shorttitle{Effect of the CMB in (sub-)mm observations}
\shortauthors{E. da Cunha et al.}

\begin{document}


\title{On the effect of the cosmic microwave background in high-redshift \\ 
(sub-)millimeter observations}


\author{Elisabete da Cunha\altaffilmark{1}, Brent Groves\altaffilmark{1}, Fabian Walter\altaffilmark{1}, Roberto Decarli\altaffilmark{1}, Axel Weiss\altaffilmark{2}, Frank Bertoldi\altaffilmark{3}, Chris Carilli\altaffilmark{4}, Emanuele Daddi\altaffilmark{5},
David Elbaz\altaffilmark{6}, Rob Ivison\altaffilmark{6,7}, Roberto Maiolino\altaffilmark{8,9}, Dominik Riechers\altaffilmark{10}, Hans-Walter Rix\altaffilmark{1},
Mark Sargent\altaffilmark{5}, Ian Smail\altaffilmark{11}}
\affil{$^1$Max-Planck-Institut f\"ur Astronomie, K\"onigstuhl 17, 69117 Heidelberg, Germany}
\affil{$^2$Max-Planck-Institut f\"ur Radioastronomie, Auf dem H\"ugel 69, 53121 Bonn, Germany}
\affil{$^3$Argelander Institute for Astronomy, University of Bonn, Auf dem H\"ugel 71, 53121 Bonn, Germany}
\affil{$^4$National Radio Astronomy Observatory, Pete V. Domenici Array Science Center, P.O. Box O, Socorro, NM, 87801, USA}
\affil{$^5$Laboratoire AIM, CEA/DSM-CNRS-Universit\'e Paris Diderot, Irfu/Service d'Astrophysique, CEA Saclay, \\
Orme des Merisiers, 91191 Gif-sur-Yvette Cedex, France}
\affil{$^6$ UK Astronomy Technology Centre, Royal Observatory, Blackford Hill, Edinburgh EH9 3HJ, United Kingdom}
\affil{$^7$ Institute for Astronomy, University of Edinburgh, Blackford Hill, Edinburgh EH9 3HJ, United Kingdom}
\affil{$^8$Cavendish Laboratory, University of Cambridge, 19 J.J. Thomson Avenue, Cambridge, CB3 0HE, United Kingdom}
\affil{$^9$Kavli Institute for Cosmology, University of Cambridge, Madingley Road, Cambridge CB3 OHA, United Kingdom}
\affil{$^{10}$Department of Astronomy, Cornell University, Ithaca, NY 14853, USA}
\affil{$^{11}$Institute for Computational Cosmology, Durham University,  South Road,  Durham DH1 3LE,  United Kingdom}

\email{cunha@mpia.de}

\begin{abstract}
\noindent
Modern (sub-)millimeter interferometers enable the measurement of the cool gas and dust emission of high-redshift galaxies ($z>5$).
However, at these redshifts the cosmic microwave background (CMB) temperature is higher, approaching, and even exceeding, the temperature of cold dust and molecular gas observed in the local Universe. In this paper, we discuss the impact of the warmer CMB on (sub-)millimeter observations of high-redshift galaxies.
The CMB affects the observed (sub-)millimeter dust continuum and the line emission (e.g. carbon monoxide, CO) in two ways: (i) it provides an additional source of (both dust and gas) heating; and (ii) it is a non-negligible background against which the line and continuum emission are measured. We show that these two competing processes affect the way we interpret the dust and gas properties of high-redshift galaxies using spectral energy distribution models. We quantify these effects and provide correction factors to compute what fraction of the intrinsic dust (and line) emission can be detected against the CMB as a function of frequency, redshift and temperature.
We discuss implications on the derived properties of high-redshift galaxies from (sub-)millimeter data. Specifically, the inferred dust and molecular gas masses can be severely underestimated for cold systems if the impact of the CMB is not properly taken into account.

\end{abstract}

\keywords{
galaxies: ISM -- galaxies: evolution -- sub-millimeter: galaxies, ISM.
}

\section{Introduction} \label{sect:intro}

Modern (sub-)millimeter interferometers allow us to routinely measure the gas and dust content of very high redshift galaxies ($z>5$), giving precious insight into the star formation properties and physical state of the interstellar medium (ISM) in early galaxies. With ALMA, for example, thanks to the combination of increased sensitivities and the negative $k$-correction in the (sub-)millimeter, it will be possible, for the first time, to detect the dust continuum and CO lines from galaxies with total luminosities that are close to that of the Milky Way. Instead of detecting only high-luminosity starbursts at high redshifts (i.e. sub-millimeter galaxies; e.g.~\citealt{Blain2002}), we will gain access to the larger population of low-luminosity (with infrared luminosities $<10^{12}~\lsun$) galaxies. In the local Universe, the bulk of dust in normal star-forming galaxies of moderate infrared luminosity typically has temperatures of $\simeq 20$~K (e.g.~\citealt{Smith2012}). However, at higher redshifts, the cosmic microwave background (CMB) temperature approaches and can even surpass this temperature. This means that we must consider the effects of the CMB on our observations.

The temperature of the CMB at any redshift $z$ is given by:
\begin{equation}
\tcmb(z)=\tcmbo\, (1+z) \,,
\label{tcmb}
\end{equation}
where \tcmbo\ is the temperature of the CMB at $z=0$, $\tcmbo=2.73$~K, sets a fundamental minimum temperature of the ISM (assuming local thermal equilibrium, LTE).
The increase of the minimum ISM temperature with redshift affects the physical conditions of the dust and molecular gas in galaxies, boosting both the dust continuum emission and the line luminosities (e.g.~\citealt{Silk1997,Blain1999,Combes1999,Righi2008}). However, at high redshift, the CMB also becomes a stronger background against which both the dust continuum and line fluxes are measured (e.g. in \citealt{Combes1999,Papadopoulos2000,Obreschkow2009,Lidz2011,Munoz2013}). These two competing effects have been discussed previously, however only in the context of CO line emission (e.g. in \citealt{Combes1999,Obreschkow2009} and briefly in other works cited above, sometimes incorrectly: \citealt{Silk1997}). A full exploration of the effects of the CMB on both continuum and line emission observations is needed now that ALMA is in operation. In this paper, we quantify in detail the implications of the CMB on the interpretation of (sub-)millimeter observations in terms of galaxy intrinsic dust and gas properties. Deriving the correct intrinsic galaxy properties for (sub-)mm observations, such as the star formation rate (via the total infrared luminosity) and the molecular gas mass (via the CO line luminosity or the dust mass) is crucial to our understanding of the
star formation efficiency of galaxies at high-redshift (e.g.~\citealt{Daddi2010b,Genzel2010,Magdis2012}).

In this study, we wish to analyze solely the effect of the CMB background at high redshift on (potential) observations of `Milky-Way-like' galaxies at high redshift. Therefore, we fix the intrinsic properties of the galaxies under consideration and assume no evolution in dust properties and stellar radiation field, in order to isolate the CMB effects. We note, however, that it is likely that high-redshift galaxies have considerably different properties than low-redshift galaxies of the same luminosity. For example, it is likely that galaxies with cold ISM such as the Milky Way do not exist due to the harder radiation fields in low metallicity environments. Also, the physical properties of dust grains may be different at high redshifts. These differences are beyond the scope of this paper.

In Section~\ref{continuum}, we describe the effects of the CMB on the dust continuum of galaxies and the implications on the derived physical properties of the dust; we also provide a recipe to take the effects of the CMB into account when comparing models with observations. In Section~\ref{lines}, we analyze the effect of the CMB on the observed CO line emission of galaxies, focusing on the simplest LTE case, and two more general examples of non-LTE cases. Our conclusions are summarized in Section~\ref{conclusion}.

\section{Effect of the CMB on (sub-)mm continuum emission}
\label{continuum}

\subsection{Dust temperature}
\label{section_temp}

We consider a galaxy at $z=0$, where diffuse dust is being heated by the radiation field produced by stars in the galaxy, and the effects of the CMB radiation are negligible. We assume that the dust is in thermal equilibrium with the radiation field, with an equilibrium temperature $\tdusto=18$~K, typical of the diffuse ISM of the Milky Way (e.g.~\citealt{daCunha2008}; see also, e.g.~\citealt{Groves2012} for an analysis of the cold dust in M31).
If we place exactly the same galaxy (i.e. same starlight intensity heating the dust, same dust properties etc.) at $z=6$, where the CMB temperature has increased from $2.73$~K to $19.1$~K, i.e. higher than \tdusto\ (eq.~\ref{tcmb}), how does this higher CMB temperature affect the dust temperature of the galaxy? The dust grains will absorb the CMB photons, so the temperature of dust in the galaxy at $z=6$ will be higher than at $z=0$ even if the other properties of the galaxy are exactly the same.
In the following, we derive the temperature of the dust in a galaxy at redshift $z$, $\tdust(z)$, when heating by the CMB is taken into account.

\begin{figure}
\begin{center}
\includegraphics[width=0.5\textwidth]{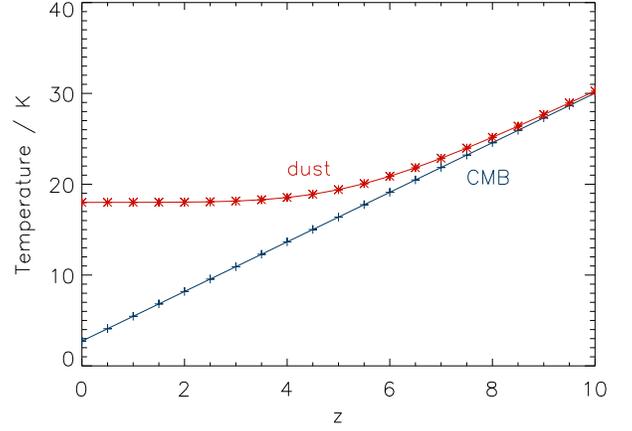}
\caption{Variation of the dust temperature (red; eq.~\ref{tdustz}) and CMB temperature (blue; eq.~\ref{tcmb}) with redshift. For dust grains in a galaxy with equilibrium temperature at $z=0$, $\tdusto=18$~K, the effect of additional dust heating by the CMB starts being non-negligible around $z\simeq4$.}
\label{tdust}
\end{center}
\end{figure}

If the dust grains are in {\em thermal equilibrium}, then they emit energy at the same rate they it absorb it, i.e.:
\begin{equation}
\frac{dE_\mathrm{em}}{dt}=\frac{dE_\mathrm{abs}}{dt}\,.
\label{energy}
\end{equation}
For a single dust grain, the energy loss rate through emission is:
\begin{equation}
\frac{dE_\mathrm{em}}{dt}=4\pi \int_0^\infty d\nu\, B_\nu[T_\mathrm{dust}(z)]\,\pi a^2\,Q_\mathrm{em}(\nu,a)\,,
\end{equation}
while the rate of energy absorbed per grain is:
\begin{equation}
\frac{dE_\mathrm{abs}}{dt}=4\pi \int_0^\infty d\nu\,\pi a^2\,Q_\mathrm{abs}(\nu,a)\,I_\nu\,,
\end{equation}
where $\tdust(z)$ is the equilibrium temperature of the dust grains at redshift $z$, and $Q_\mathrm{em}(\nu,a)$ and $Q_\mathrm{abs}(\nu,a)$ are, respectively, the emission and absorption coefficients at frequency $\nu$, for a grain of effective radius $a$ (see, e.g., \citealt{Spitzer1978,Draine2011} for more detail). In general, $Q_\mathrm{em}(\nu,a)$=$Q_\mathrm{abs}(\nu,a)$ \citep{Draine1984}, and we assume that the dust properties are invariant with redshift, i.e. that $Q_\mathrm{abs}(\nu,a)$ does not change with $z$.

At any redshift, we can write the intensity of the radiation field heating the dust as (e.g.~\citealt{Rowan1979}):
\begin{equation}
I_\nu(z)=J^\ast_\nu(z) + B_\nu[\tcmb(z)] \,,
\label{inu}
\end{equation}
where the first term, $J^\ast_\nu(z)$, corresponds to the contribution by the radiation produced by stars in the galaxy, and the second term, $B_\nu[\tcmb(z)]$, is the contribution by the CMB radiation, which emits as a black body $B_\nu$ of temperature $\tcmb(z)$. Thus, at a given redshift, using eq.~\ref{inu}, the equation of the energy balance of the dust grains in thermal equilibrium (eq.~\ref{energy}) takes the form:
\begin{align}
\int_0^\infty d\nu\ Q_\mathrm{abs}(\nu,a) \, J^\ast_\nu(z)  + \int_0^\infty d\nu\ Q_\mathrm{abs}(\nu,a) \, B_\nu[\tcmb(z)] \nonumber \\
= \int_0^\infty d\nu\ Q_\mathrm{abs}(\nu,a) \, B_\nu[\tdust(z)]\,.
\end{align}
From this equation, the total energy produced by stars in the galaxy (that is absorbed by the dust grains) can be written as:
\begin{align}\label{energy3}
\int_0^\infty d\nu\ Q_\mathrm{abs}(\nu,a) \, J^\ast_\nu(z) \nonumber \\
= \int_0^\infty d\nu\ Q_\mathrm{abs}(\nu,a) \, \{ B_\nu[\tdust(z)] - B_\nu[\tcmb(z)] \} \,.
\end{align}
We assume here that the properties of the galaxy do not change with redshift, i.e. the stellar radiation field, $J^\ast_\nu$, remains the same. Therefore, for any redshift $z$,
\begin{equation}
\int_0^\infty d\nu\ Q_\mathrm{abs}(\nu,a) \, J^\ast_\nu(z) = \int_0^\infty d\nu\ Q_\mathrm{abs}(\nu,a) \, J^\ast_\nu(z=0) \,,
\label{energy4}
\end{equation}
and combining with eq.~\ref{energy3} we can write:
\begin{align}\label{energy5}
\int_0^\infty d\nu\ Q_\mathrm{abs}(\nu,a) \, \{ B_\nu[\tdust(z)] - B_\nu[\tcmb(z)] \} \nonumber \\
= \int_0^\infty d\nu\ Q_\mathrm{abs}(\nu,a) \, \{ B_\nu[\tdusto] - B_\nu[\tcmbo] \} \,.
\end{align}

At (sub-)mm wavelengths, the grain emissivity can be approximated with a power-law function with frequency, $Q_\mathrm{abs}(\nu,a) \propto \nu^\beta$, where $\beta$ is the so-called dust emissivity index (e.g.~\citealt{Draine1984}). Thus, we can replace this term in eq.~\ref{energy5} and re-arrange in order to solve for $\tdust(z)$ as follows:
\begin{align}\label{energy6}
\int_0^\infty d\nu\ \nu^\beta B_\nu[\tdust(z)] & = \int_0^\infty d\nu\ \nu^\beta B_\nu[\tcmb(z)] \nonumber \\
& + \int_0^\infty d\nu\ \nu^\beta B_\nu[\tdusto]  \nonumber \\
& - \int_0^\infty d\nu\ \nu^\beta B_\nu[\tcmbo] \,.
\end{align}
The integral of a modified black body $\nu^\beta B_\nu(T)$ is:
\begin{equation}
\int_0^\infty d\nu\ \nu^\beta B_\nu(T) \propto T^{4+\beta} \,.
\label{energy7}
\end{equation}
From eqs.~\ref{tcmb}, \ref{energy6} and \ref{energy7}, we obtain the following equation for the equilibrium dust temperature at redshift $z$:
\begin{equation}
\tdust(z) = \Big( (\tdusto)^{4+\beta} + (\tcmbo)^{4+\beta} \big[(1+z)^{4+\beta} - 1 \big] \Big) ^{\frac{1}{4+\beta}}\,.
\label{tdustz}
\end{equation}

In Fig.~\ref{tdust}, we plot the evolution of dust and CMB temperatures with redshift, assuming a dust temperature of 18~K at $z=0$. This shows that, for this temperature, the effect of dust heating by the CMB becomes non-negligible at $z\simeq4$; the higher the redshift, the dust temperature asymptotically approaches the temperature of the CMB.

\subsection{Effect on the intrinsic far-IR/sub-mm dust SEDs}
\label{section_seds}

\begin{figure*}
\begin{center}
\includegraphics[width=0.95\textwidth]{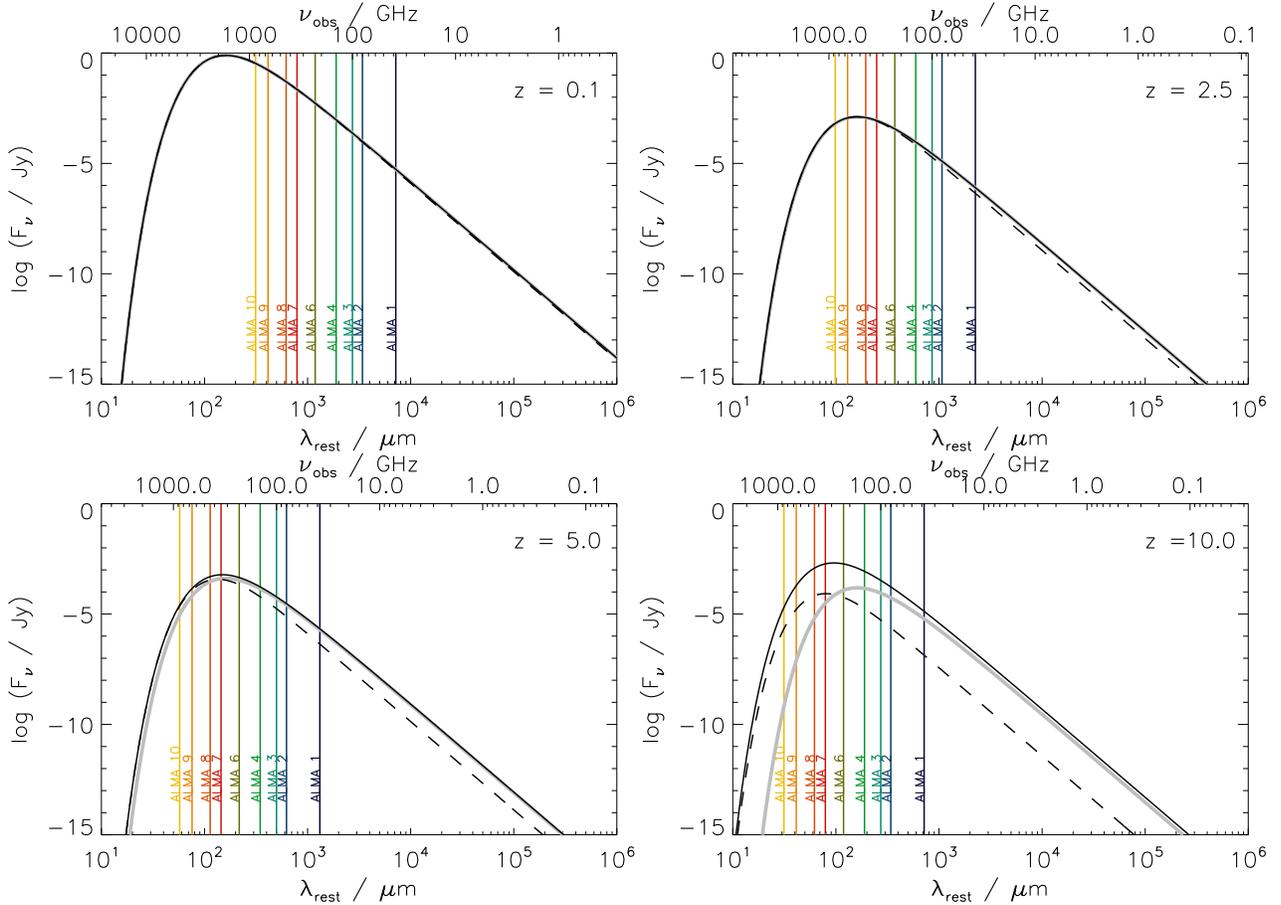}
\caption{Spectral energy distributions of dust grains in thermal equilibrium with $\tdusto=18$~K, with total luminosity powered by starlight in the galaxy of $10^{11}~\lsun$, at redshifts $z=0.1, 2.5, 5$~and $10$ (redshift is indicated in the upper-right corner of each plot). {\em grey line} -- intrinsic SED when no dust heating by the CMB is included; {\em black solid line} -- intrinsic SED including dust heating by the CMB; {\em black dashed line} -- actual measured SED, when contrast with the CMB background is included (Section~\ref{cmb_contrast}). The colored vertical lines indicate the observed frequency (and rest-frame wavelength) sampled by the 9 ALMA bands ($\nu_\mathrm{obs}=38$, 80, 100, 144, 230, 345, 430, 660 and 870~GHz).}
\label{seds}
\end{center}
\end{figure*}

Fig.~\ref{tdust} shows that the effect of dust heating by the CMB on the (cold) dust temperature of galaxies is to first order negligible at redshifts $z\lesssim 4$. However, beyond this redshift, the dust temperature increases with redshift due to the extra heating by the CMB as demonstrated in the previous section. To investigate how this affects the far-infrared and sub-millimeter SEDs of the galaxies, we plot, in Fig.~\ref{seds}, the predicted intrinsic SED of dust grains in thermal equilibrium with $\tdusto=18$~K, at redshifts $z=0.1, 2.5, 5$~and $10$. In each panel, the thick grey line is identical and shows the un-corrected SED (dust heating only by star formation i.e. the radiation field $J^\ast_\nu$) and the black line shows the predicted intrinsic SED when taking into account additional dust heating by the CMB. We stress that this is the intrinsic SED, not the actually observed one (as discussed in Section~\ref{cmb_contrast}). As the dust temperature increases due to increasing CMB temperature, this affects the dust SEDs in two ways, as seen in Fig.~\ref{seds}: (i) the peak of the SED shifts towards lower (rest-frame) wavelengths (i.e., higher observed frequencies) as the dust temperature increases with redshift; (ii) the total luminosity increases by a factor $[\tdust(z)/\tdusto]^{(4+\beta)}$ (eq.~\ref{energy7}), due to additional energy from CMB photons absorbed by the dust grains.

\begin{figure}
\begin{minipage}{\linewidth}
\centering
\includegraphics[width=\textwidth]{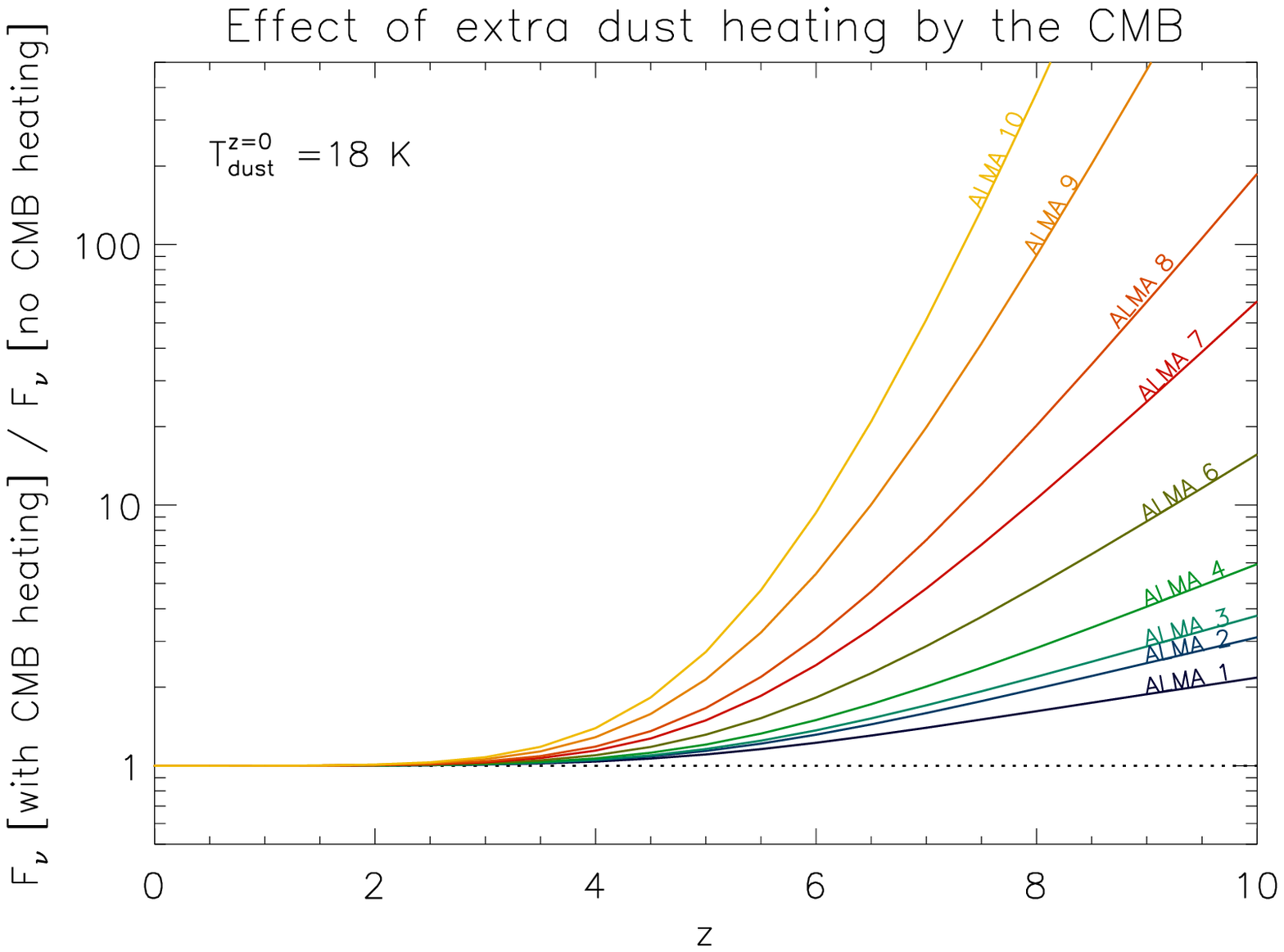}
\end{minipage}
\begin{minipage}{\linewidth}
\centering
\includegraphics[width=\textwidth]{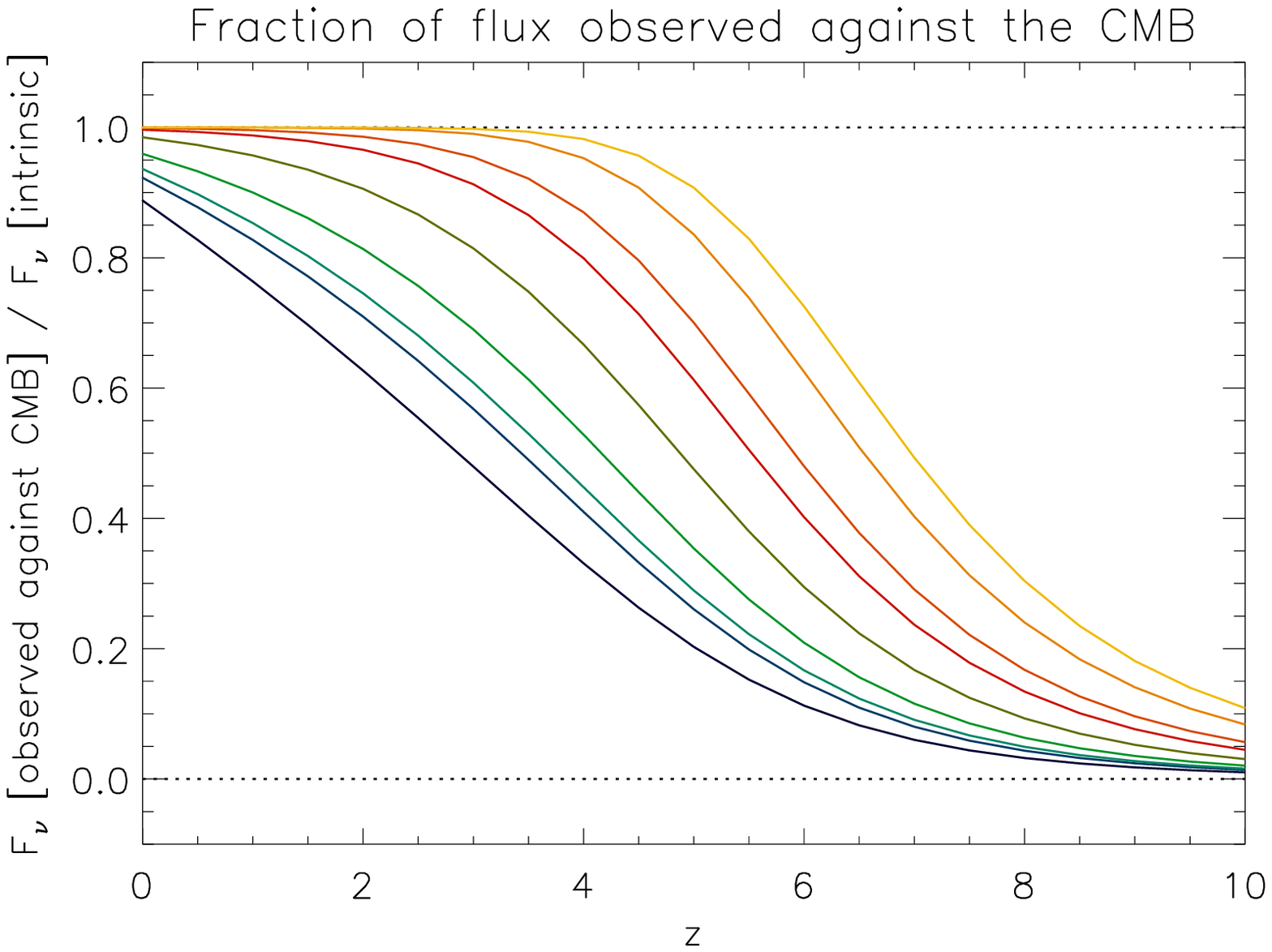}
\end{minipage}
\caption{{\em Top panel:} Effect of additional dust heating by the increasing CMB temperature with redshift. We plot the ratio between the predicted flux density in each ALMA band when including dust heating by the CMB and not including dust heating by the CMB, for dust grains with an equilibrium temperature of 18~K at $z=0$. This figure does not include contrast with the CMB background. This shows, as in Fig.~\ref{seds}, that the highest frequency bands are the most affected by dust heating by the CMB. {\em Bottom panel:} Effect of the CMB as an observing background. We plot the ratio between the flux in each ALMA band that can be measured against the CMB at a given redshift, and the intrinsic flux emitted by the galaxy at that frequency. As for the top panel, we assume $\tdusto=18$~K and, at each redshift, include the extra heating contributed by the CMB in the intrinsic emitted flux.}
\label{frac_flux}
\end{figure}
Finally, to check the impact on the intrinsic (sub-)mm continuum flux densities, in the top panel of Fig.~\ref{frac_flux} we plot the effect of the CMB heating on the continuum fluxes in 9 ALMA bands from 38 to 870~GHz, as a function of redshift. This shows that the CMB has practically no effect on the fluxes up to $z\simeq4$, but after that redshift it contributes increasingly to boost the intrinsic (sub-)mm fluxes in the ALMA bands. The effect is strongest for the higher-frequency bands, which sample the dust SEDs closer to the peak (see Fig.~\ref{seds}).

\subsection{Detectability of dust emission against the CMB background}
\label{cmb_contrast}

Extra dust heating as described in Section~\ref{section_seds} is not the only effect of the CMB radiation on the observed (sub-)millimeter fluxes. For any (sub-)mm galaxy observation, the dust continuum is always measured against the CMB. Here we discuss how this affects the detectability of the dust continuum of galaxies at high redshifts.

We assume that the stellar radiation can be neglected at (sub-)millimeter wavelengths, i.e. $J^\ast_\nu=0$. From solving the radiative transfer equation, the (rest-frame) intensity per unit frequency from a galaxy at redshift $z$ in the (sub-)millimeter is:
\begin{equation}
I_\nu= [1-\exp({-\tau_\nu})]\,B_\nu[T_\mathrm{dust}(z)] + \exp({-\tau_\nu})\,B_\nu[T_\mathrm{CMB}(z)]\,,
\label{inu_rest}
\end{equation}
where $\tau_\nu$ is the dust optical depth. The first term of this equation corresponds to the emission by dust in thermal equilibrium with temperature $T_\mathrm{dust}(z)$ (given by eq.~\ref{tdustz}), and the second term is the contribution from CMB radiation that is transmitted through the galaxy ISM (i.e. the fraction of the underlying CMB that is not absorbed by dust).
We assume that the dust is optically thin in the (sub-)millimeter\footnote{This is a reasonable assumption, as the dust column required for the galaxy to be optically thick at (sub-)millimeter wavelengths are unrealistically high. For example, using a dust mass absorption coefficient of $\kappa_\nu(850~\mathrm{\mu m})=0.77$~g$^{-1}$~cm$^2$ \citep{Dunne2000}, a galaxy of total dust mass $10^8$~\msun\ would need to have a radius smaller than 100~pc to be optically thick at 850~\mic.}, i.e. $\tau_\nu \ll 1$, then $\exp(-\tau_\nu) \approx 1-\tau_\nu$. The optical depth can be expressed as: $\tau_\nu=\Sigma_d \kappa_\nu$, where $\Sigma_d$ is the surface mass density of dust\footnote{This can be written in terms of the total dust mass $M_d$ and the physical area of the galaxy $A$: $\Sigma_d=M_d/A$.} (in units of g~cm$^{-2}$) and $\kappa_\nu$ is the mass absorption coefficient (cross-section per unit mass, in units of g$^{-1}$ cm$^{-2}$). Therefore, eq.~\ref{inu_rest} can be re-written as:
\begin{equation}
I_\nu =\Sigma_d \kappa_\nu \,B_\nu[T_\mathrm{dust}(z)] + (1-\Sigma_d \kappa_\nu)\,B_\nu[T_\mathrm{CMB}(z)]\,.
\label{inu_rest2}
\end{equation}
The observed flux of the galaxy is:
\begin{equation}
F_{\nu/(1+z)}^\mathrm{obs}=\Omega\frac{I_\nu}{(1+z)^3}\,,
\label{fnu_obs}
\end{equation}
where $\Omega$ is the solid angle subtended by the galaxy, which can be written in terms of the physical area of the galaxy $A$ and the luminosity distance $d_\mathrm{L}$, as: $\Omega=(1+z)^4\,A/d_\mathrm{L}^2$.
From eqs.~\ref{inu_rest2} and \ref{fnu_obs} we obtain:
\begin{equation}
F_{\nu/(1+z)}^\mathrm{obs} = \frac{\Omega}{(1+z)^3} \Big[\Sigma_d \kappa_\nu \,B_\nu[T_\mathrm{dust}(z)] + (1-\Sigma_d \kappa_\nu)\,B_\nu[T_\mathrm{CMB}(z)]\Big]\,.
\label{fnu_obs2}
\end{equation}

The flux is always measured relative to the CMB background. The flux measured against the CMB is obtained by subtracting the observed CMB flux (over the solid angle $\Omega$) from the predicted flux in eq.~\ref{fnu_obs2}. We obtain\footnote{The observed CMB intensity is related to the rest-frame CMB intensity as: $B_{\nu/(1+z)}(\tcmbo)=B_\nu[T_\mathrm{CMB}(z)] / (1+z)^3$.}:
\begin{align}
F_{\nu/(1+z)}^\mathrm{obs\,against\,CMB} & =  F_{\nu/(1+z)}^\mathrm{obs} - \Omega\, B_{\nu/(1+z)}\big[\tcmbo\big]  \nonumber \\
 & = F_{\nu/(1+z)}^\mathrm{obs} - \Omega\, \frac{B_\nu\big[T_\mathrm{CMB}(z)\big]}{(1+z)^3}   \nonumber \\
 & = \frac{\Omega}{(1+z)^3} \Sigma_d \kappa_\nu \Big( B_\nu\big[T_\mathrm{dust}(z)\big] - B_\nu\big[T_\mathrm{CMB}(z)\big] \Big) \nonumber \\
 & = \frac{1+z}{d_\mathrm{L}^2} M_d \kappa_\nu \Big( B_\nu\big[T_\mathrm{dust}(z)\big] - B_\nu\big[T_\mathrm{CMB}(z)\big] \Big)\,.
\label{fnu_cmb1}
\end{align}

\begin{figure*}[t]
\begin{center}
\includegraphics[width=\textwidth]{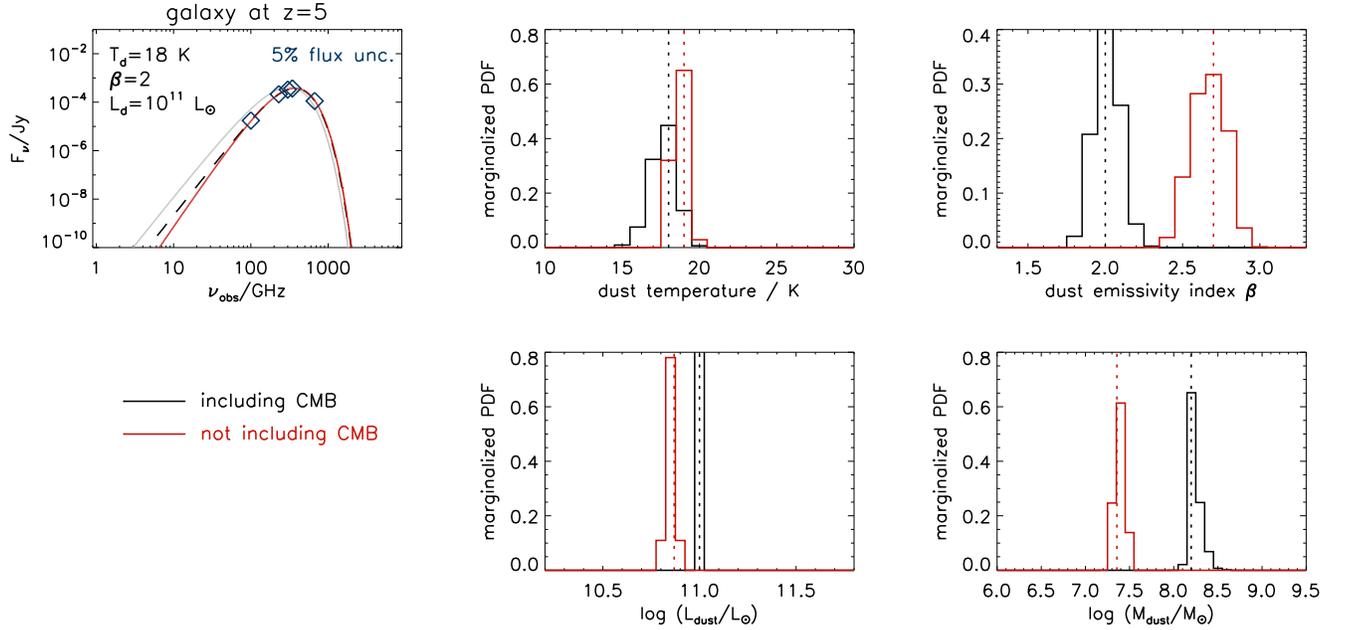}
\caption{Comparison between fits to our fiducial model SED (intrinsic dust temperature of $\tdusto=18$~K, dust emissivity index $\beta=2$, total luminosity $10^{11}$~\lsun, and total dust mass $1.6\times10^8$~\msun) at $z=5$, when including the effect of the CMB in the fit (in black) and when ignoring the effect of the CMB in the fit (in red). The `observed' fluxes are indicated as blue diamonds in the top-left panel; we use five common (sub-)millimeter bands at 100, 230, 300, 345 and 670~GHz. The grey SED is the intrinsic SED, i.e. before CMB corrections -- the actual observed SED i.e. with the effects of the CMB included as for Fig.~\ref{seds}, is plotted as a dashed black line. The best-fit SED when fitting the blue fluxes but ignoring the effects of the CMB is plotted in red. The top-right panels and the two bottom panels show the probability density functions (PDFs) of the constrained parameters (dust temperature, emissivity index, luminosity and mass) computed using a Bayesian fitting method \citep{daCunha2008}, with the best-fit value for each parameter indicated by a vertical dashed line.}
\label{sed_fit1}
\end{center}
\end{figure*}

Based on this equation, the fraction of the intrinsic dust emission from the galaxy (i.e. the first term of eq.~\ref{inu_rest}) that we can actually measure against the CMB at a given frequency $\nu_\mathrm{obs}=\nu/(1+z)$ is given by:
\begin{align}
\frac{F_{\nu/(1+z)}^\mathrm{obs\,against\,CMB}}{F_{\nu/(1+z)}^\mathrm{intrinsic}}  & =\frac{B_\nu\big[T_\mathrm{dust}(z)\big] -B_\nu\big[T_\mathrm{CMB}(z)\big]}{B_\nu\big[T_\mathrm{dust}(z)\big] } \nonumber \\
& = 1 - \frac{B_\nu\big[T_\mathrm{CMB}(z)\big] }{B_\nu\big[T_\mathrm{dust}(z)\big] }\,.
\label{continuum_contrast}
\end{align}
If $T_\mathrm{dust}(z)=T_\mathrm{CMB}(z)$, i.e. the dust is in thermal equilibrium with the CMB, then no intrinsic flux can be detected against the CMB; on the other hand, if $T_\mathrm{dust}(z) \gg T_\mathrm{CMB}(z)$ (usually the case at low-$z$), then practically all the intrinsic flux is detected against the CMB. In the bottom panel of Fig.~\ref{frac_flux}, we plot the fraction of flux observed against the CMB as a function of redshift in the 9 ALMA bands from 38 to 870~GHz, for a galaxy with intrinsic dust temperature $\tdusto=18$~K. This figure shows that, as the redshift increases and the CMB temperature gets closer to the dust temperature (as shown in Fig.~\ref{tdust}), it becomes increasingly difficult to detect the dust emission against the CMB, and this effect is higher for the lower-frequency bands. In Fig.~\ref{seds}, we add a black dashed line that represents the actually {\em observed} SED of a galaxy with dust intrinsic temperature $\tdusto=18$~K at different redshifts. This is obtained by multiplying the intrinsic SED emitted at each redshift (including heating by the CMB; black solid lines in Fig.~\ref{seds}) by the factor given by eq.~\ref{continuum_contrast} at each redshift. Figs.~\ref{seds} and \ref{frac_flux} (bottom panel) show that the fraction of flux from a galaxy at a given redshift that can be detected against the CMB varies with frequency. This can have severe implications on the interpretation of observed (sub-)millimeter SEDs at high redshift in terms of dust properties. For example, for the SED in Fig.~\ref{seds}, at $z=5$, the Rayleigh-Jeans tail of the dust emission observed against the CMB (black dashed line) is much steeper than in the intrinsically emitted SED (black solid line), which would be interpreted as a higher emissivity index $\beta$ of the dust grains, and would change the inferred dust luminosity, mass and temperature, as we discuss in more detail in the next section.

\subsection{Discussion: effect on the physical interpretation of the observed (sub-)mm dust emission}
\label{discussion_dust}

\begin{figure*}
\begin{center}
\includegraphics[width=0.85\textwidth]{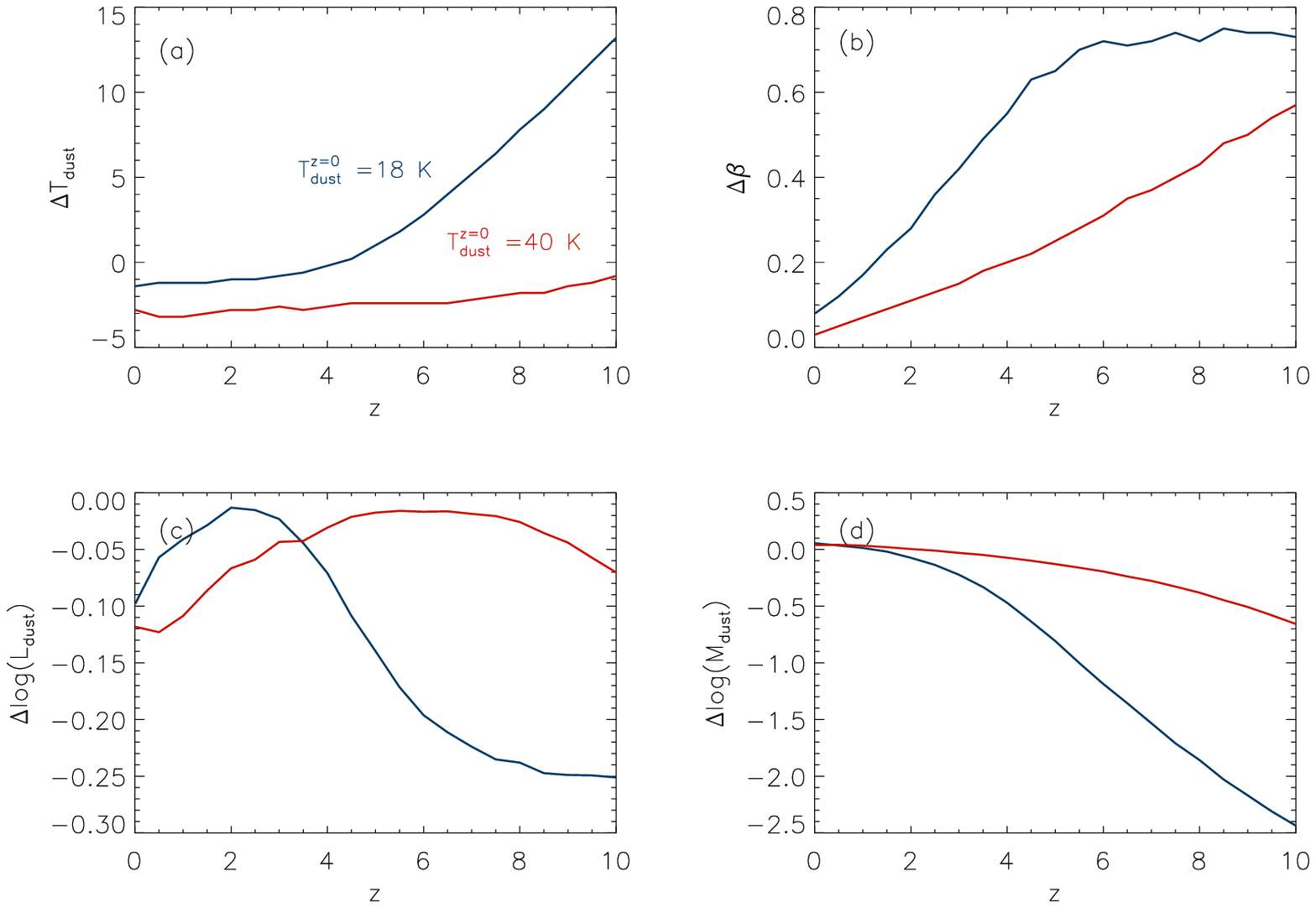}
\caption{Difference between different dust parameter estimates when not including the CMB effects minus when including the CMB in the fits (fitting the fluxes in five bands with $\nu_\mathrm{obs}=100, 230, 300, 345, 670$~GHz), for the dust emission model with $\beta=2$ and intrinsic temperature $\tdusto=18$~K (in blue) and $\tdusto=40$~K (in red): (a) temperature; (b) emissivity index; (c) logarithm of the total luminosity; (d) logarithm of the total mass. This shows that, as in Fig.~\ref{sed_fit1}, the dust temperature and emissivity index tend to be overestimated when ignoring the effect of the CMB on the SEDs at high redshift, and the total dust luminosity and mass are underestimated. As expected, these systematic effects are less severe for higher intrinsic dust temperatures.} 
\label{params}
\end{center}
\end{figure*}

In the previous sections we show that the CMB has an impact on the peak of the observed dust SED and its slope in the Rayleigh-Jeans regime. Here we investigate how this may affect the inferred dust temperature, luminosity, emissivity index and mass from typical (sub-)millimeter observations.
It is usual practice to interpret observed far-infrared/sub-mm galaxy SEDs by comparing them with modified black body (MBB) functions (of the form $\nu^\beta\,B_\nu[T_\mathrm{dust}]$) that describe the emission by dust in thermal equilibrium using only three parameters: the dust emissivity index $\beta$, temperature $T_\mathrm{dust}$, and total luminosity (e.g.~\citealt{Dunne2000,Klaas2001}).

We consider the fiducial model discussed in the previous sections, with an intrinsic dust temperature of $\tdusto=18$~K, dust emissivity index $\beta=2$, total luminosity $10^{11}$~\lsun, and total dust mass $1.6\times10^{8}$~\msun. We assume that our model galaxy is at $z=5$, and compute the predicted fluxes in five typical (sub-)millimeter bands at 100~GHz (3~mm), 230~GHz (1.3~mm), 300~GHz (1~mm), 345~GHz (870~\mic) and 670~GHz (450~\mic). These fluxes are supposed to be the `real' observed fluxes, therefore we take into account the effect of extra heating and extra background provided by the CMB at $z=5$, as discussed in Sections~\ref{section_temp} to \ref{cmb_contrast}. We assume a typical flux uncertainty of 5\% in each band. In order to test how biased our temperature, emissivity, luminosity and dust mass estimates for this model galaxy would be if we ignored the effect of the CMB, we fit the model fluxes in the five (sub-)millimeter bands using a set of models that does not include the effect of the CMB; we also compare with a set of models that correctly includes the effect of the CMB. In practice, we compare our synthetic observed fluxes in five bands to a grid of modified black bodies where we vary the intrinsic dust temperature \tdusto\ between 10 and 30~K, and the emissivity index $\beta$ between 1 and 3.

In Fig.~\ref{sed_fit1}, we plot the results of this fitting when including the CMB effects (in black) and the results when ignoring the additional effects of the CMB (in red); for each case, we plot  the best-fit SED and the marginalized probability distribution functions (PDFs; computed using the approach described in \citealt{daCunha2008}) for the dust temperature, emissivity, total luminosity and mass, with the best-fit value for each parameter (i.e. the value that minimizes $\chi^2$) indicated by a vertical dashed line. We first check that the best-fit model parameters obtained when including the CMB effect (indicated by the black dotted lines) are equal to the input parameters -- this confirms the robustness of the method. When the SED is fitted with models that do not include the effects of the CMB, there is a significant difference between the best-fit model parameters and the input (i.e.~`real') parameters. The intrinsic dust temperature and emissivity index are overestimated ($\tdusto=19$~K instead of $18$~K, and $\beta=2.7$ instead of $2.0$), while the dust luminosity and the dust mass are underestimated ($\log(\ldust/\lsun)=10.87$ instead of $11.00$, and $\log(\mdust/\msun)=7.35$ instead of $8.20$). The effect on the estimated dust mass is the strongest because the dust mass depends on the luminosity, temperature and emissivity index as $M_\mathrm{dust} \propto \ldust\ T_\mathrm{dust}^{-(4+\beta)}$. This confirms what was already hinted in Section~\ref{cmb_contrast}: the CMB makes the SEDs look hotter and steeper, and Fig.~\ref{sed_fit1} shows that this has a significant impact in the deduced properties of dust in high-redshift galaxies when fitting standard MBB models to the observed dust emission. A steeper emissivity index would imply different dust properties at high redshift, and the difference in inferred dust masses would also change our understanding of ISM enrichment by dust in high-redshift galaxies.

In Fig.~\ref{params}, we investigate these effects in a more systematic way. We use the same model and perform the similar modified black body fits as in the test described in Fig.~\ref{sed_fit1}, but at various redshifts between $z=0$ and $z=10$, and for two intrinsic input dust temperatures, $\tdusto=18$~K (in blue) and $\tdusto=40$~K (in red). The four panels of Fig.~\ref{params} show the difference between the best-fit parameter value when ignoring the CMB effects on the SEDs and when including the CMB effects, for the dust temperature (a), the dust emissivity index (b), the total luminosity (c), and the dust mass (d). We show, for each temperature, the result of fitting the same five (sub-)mm bands considered in Fig.~\ref{sed_fit1}: observed frequencies $\nu_\mathrm{obs}=100$, $230$, $300$, $345$ and $670$~GHz.
Not surprisingly, the overestimation of temperature and emissivity index and the underestimation of luminosity and mass are worse at high redshifts, and for the lowest intrinsic dust temperature, since in this case the temperature contrast between the CMB and the dust emission is small. The cold dust mass can be severely underestimated, up to about two orders of magnitude at $z\simeq10$. We note that, at very low redshifts, we underestimate the dust temperature and luminosity, and overestimate the emissivity index, even though we expect the CMB to have a minimal effect at low redshifts: the differences should be close to zero in all panels at $z=0$. This offset is explained by the fact that, when fitting fluxes at {\em observed-frame} frequencies 670~GHz and lower, as is the case for the solid lines in these plots, we sample the SED significantly lower frequencies from its peak, and relatively close to the peak of the CMB emission at low redshifts.

\subsection{How to account for the CMB when interpreting real continuum measurements}

We have shown in the previous section that, when interpreting real (sub-)mm continuum measurements in terms of intrinsic dust properties, we must take the effect of the CMB into account.
This is the case for comparing observations with dust emission models, such as modified black bodies (as described above), or even more complex dust emission models (e.g.~\citealt{daCunha2008,Draine2007}). To summarize, we can estimate the intrinsic properties of the dust in an observed galaxy at redshift $z$ by following these steps:

\begin{enumerate}
\item for a given model with intrinsic temperature \tdusto\ and emissivity index $\beta$, compute the extra heating provided by the CMB using eq.~\ref{tdustz}; the actual dust temperature is then $\tdust(z)$;
\item include the extra luminosity provided by the CMB heating, i.e. multiply the dust emission model by $[\tdust(z)/\tdusto]^{(4+\beta)}$;
\item compute dust emission model in the observed frame, $F_{\nu/(1+z)}^\mathrm{intrinsic}$;
\item to account for the effect of the CMB as an observing background, obtain $F_{\nu/(1+z)}^\mathrm{obs\,against\,CMB}$ by multiplying the dust emission model by the factor given in eq.~\ref{continuum_contrast}: $1 - \{ B_\nu[T_\mathrm{CMB}(z)] / B_\nu[T_\mathrm{dust}(z)] \} $;
\item compare the modified dust emission model (parameterized in terms of the intrinsic dust properties from step 1) directly with the observations.
\end{enumerate}

\section{Effect of the CMB on the CO line emission}
\label{lines}

We now discuss the effect of the CMB on the CO emission at high redshifts. The mechanisms by which the CMB affects the CO excitation are not as straightforward as for the dust case discussed in the previous section, since the local thermal equilibrium (LTE) conditions can only be assumed in particular cases, and in reality molecular clouds are often in non-LTE conditions.
In general, the effect of the CMB on the detectability of high-redshift CO lines is two-fold, as in the case of dust emission. On one hand, the higher temperature of the CMB at higher redshifts can affect the excitation of CO by helping populate high rotational levels (through mechanisms that we discuss below), thus increasing the line luminosities for high-rotational number transitions (e.g.~\citealt{Silk1997}). 
On the other hand, as the CMB temperature increases, it becomes a more important observing background against which the CO lines must be detected (see also~\citealt{Combes1999,Papadopoulos2000,Obreschkow2009}).

For simplicity we assume throughout this section that the gas and dust are efficiently coupled, such that the dust temperature, \tdust, sets the minimum kinetic temperature of the gas, \tkin. This implies that the gas heating via collisions with dust grains is 100\% effective \citep{Tielens2005}. This can be the case for high optical depths \citep{Tielens1985}, but, in most realistic cases, the heating of gas by dust may not be 100\% effective. However, even in less optimal conditions the gas and dust temperatures are coupled so that if the dust temperature increases, the gas temperature increases by a similar amount \citep{Tielens2005}. We discuss deviations from the assumption that $\tkin=\tdust$ where relevant.

\subsection{Some general aspects of CO excitation}
\label{co_general}

Here we summarize some of the main aspects of CO excitation; more details can be found e.g. in \cite{Spitzer1978} or \cite{vanderTak2007}. The frequency of the photon emitted during the rotational transition of the CO molecule from a level $J_u$ to a lower-excitation level $J_l=J_u -1$ is: $\nu_{ul} = [ E(J_u) - E(J_l) ] / h = h J_u / (4\pi^2 m r_e^2) \simeq \nuco\,J_u$, where $h$ is the Planck constant, $m$ is the reduced mass of the CO molecule, $r_e$ is the effective distance between the two atoms, and $\nuco=115.2712$~GHz\footnote{In reality, the frequency of $J_u > 1$ transitions are close but not exactly multiples of \nuco\ because rotation of the molecule makes $r_e$ vary slightly (due to centrifugal forces); the exact frequencies of the transitions can be found in http://splatalogue.net.}.

The transition between rotational energy levels of CO in molecular clouds occurs mainly via collisions with H$_2$ molecules, with rates given by the collision coefficients $C_{ul}$ (for collisional de-excitation) and $C_{lu}$ for (collisional excitation). These collision coefficients depend mainly on the velocity-integrated collision cross-section (which depends among other things on the kinetic temperature of the gas, $T_\mathrm{kin}$), and the number density of collision partners, in this case $n_\mathrm{H_2}$.
Photons of frequency $\nu_{ul}$ are emitted when a molecule goes from an upper energy level $J_u$ level to a lower energy level $J_l$  via spontaneous emission, which occurs at a rate given by the Einstein coefficient $A_{ul}$, or by stimulated emission, which occurs at a rate $B_{ul} \tilde U$, where $B_{ul}$ is the Einstein stimulated emission coefficient and $\tilde U$ is the line profile-weighted mean energy density of the radiation field. Photons of the same frequency from the radiation field are absorbed at a rate $B_{lu} \tilde U$, where $B_{lu}$ is the Einstein absorption coefficient.
The Einstein probability coefficients are related by:
\begin{equation}
g_l B_{lu} = g_u B_{ul} = \frac{c^3}{8\pi h \nu_{ul}^3} g_u A_{ul} \,, 
\end{equation}
where $g_u$ and $g_l$ are the statistical weights of the levels (given by $2J_u+1$ and $2J_l+1=2J_u-1$, respectively).
The excitation state and line absorption/emission by CO molecules is computed by taking into account collisions and radiative transitions.
In statistical equilibrium, the populations of levels $u$ and $l$ are given by:
\begin{equation}
n_u (A_{ul} + B_{ul} \tilde U + C_{ul} ) = n_l (B_{lu} \tilde U + C_{lu} ) \,,
\label{stateq}
\end{equation}
where $\tilde U = 4\pi \tilde I / c$ is the average energy density of the radiation field, and $\tilde I$ is the intensity of the radiation field $I_\nu$ averaged over the line profile. We can assume $I_\nu=B_\nu(T_\mathrm{rad})$, where $T_\mathrm{rad}$ is the radiation temperature.

The `excitation temperature' of a transition from $J_u$ to $J_l$, \tex, by definition, characterizes the level populations resulting from eq.~\ref{stateq} using the Boltzmann equation:
\begin{equation}
\frac{n_u}{n_l}=\frac{g_u}{g_l}\exp\Big(-\frac{h\nu}{k \tex}\Big)\,.
\label{texc}
\end{equation}

If collisions dominate the excitation of CO, then the level populations are set by the collision coefficients and the kinetic temperature, i.e.
\begin{equation}
\frac{C_{lu}}{C_{ul}}=\frac{n_u}{n_l}=\frac{g_u}{g_l}\exp\Big(-\frac{h\nu}{k T_\mathrm{kin}}\Big)\,.
\label{tkin}
\end{equation}

The radiative transfer equation is:
\begin{equation}
\frac{dI_\nu}{ds}=-\kappa_\nu I_\nu + \epsilon_\nu \,,
\label{rt1}
\end{equation}
where $\kappa_\nu = (h\nu/c) \big( n_l B_{lu} - n_u B_{ul} \big)$ is the absorption coefficient,
and $ \epsilon_\nu= (h\nu / 4\pi) n_u A_{ul}$ is the emissivity.
The optical depth of a transition is defined as:
\begin{equation}
\tau_\nu^{J_u}=\int \kappa_\nu ds \,,
\label{tau1}
\end{equation}
and we can define the source function $\mathcal{S}_\nu$ as:
\begin{equation}
\mathcal{S}_\nu=\frac{\epsilon_\nu}{\kappa_\nu}=\frac{c}{4\pi}\frac{n_u A_{ul}}{n_l B_{lu} - n_u B_{ul}} \,,
\label{snu1}
\end{equation}
and re-write the radiative transfer equation (eq.~\ref{rt1}) as:
\begin{equation}
\frac{dI_\nu}{d\tau_\nu}=-I_\nu+\mathcal{S}_\nu \,,
\label{rtline}
\end{equation}
which has the solution:
\begin{equation}
I_\nu=\big[1-\exp\big(-\tau_\nu^{J_u}\big)\big] \mathcal{S}_\nu + I_\nu^0 \exp\big(-\tau_\nu^{J_u}\big) \,.
\label{inuline}
\end{equation}
The first term is the `intrinsic' line flux, and the second term is the transmitted background intensity. From eqs.~\ref{stateq}, \ref{texc}, and \ref{snu1} follows that:
\begin{equation}
\mathcal{S}_{\nu}=\frac{2h\nu^3}{c^2}\frac{1}{\exp\Big(\frac{h\nu}{kT_\mathrm{exc}^{J_u}}\Big)-1} = B_\nu\big(\tex\big) \,.
\label{snu2}
\end{equation}
If we assume the only background is the CMB, then the (rest-frame) intensity is:
\begin{equation}
I_\nu=\big[1-\exp\big(-\tau_\nu^{J_u}\big)\big] B_\nu\big(T_\mathrm{exc}^{J_u}\big) + \exp\big(-\tau_\nu^{J_u}\big) B_\nu\big[T_\mathrm{CMB}(z)\big]\,,
\label{inu_line}
\end{equation}
and, similarly to eq.~\ref{fnu_obs}, the observed velocity-integrated flux of the line is:
\begin{equation}
S_{\nu/(1+z)}^{J_u\mathrm{[obs]}} = \frac{\Omega}{(1+z)^3} I_\nu \,.
\label{fnu_obs_line}
\end{equation}
The velocity-integrated flux of the line measured against the CMB background is then (see also eq.~\ref{fnu_cmb1}):
\small\begin{align}
S_{\nu/(1+z)}^{J_u\mathrm{[obs\,against\,CMB]}} & = S_{\nu/(1+z)}^{J_u\mathrm{[obs]}} - \Omega B_{\nu/(1+z)} \big[\tcmbo\big] \nonumber \\
& = \frac{\Omega}{(1+z)^3} \big[1-\exp\big(-\tau_\nu^{J_u}\big)\big] \Big( B_\nu\big[\tex\big] - B_\nu\big[\tcmb(z)\big] \Big) \,.
\end{align}
\normalsize
The fraction of intrinsic line flux (first term in eq.~\ref{inu_line}) that is observed against the CMB is:
\begin{align}
\frac{S_{\nu/(1+z)}^{J_u\mathrm{[obs\,against\,CMB]}}}{S_{\nu/(1+z)}^{J_u\mathrm{[intrinsic]}}}  & = 1 - \frac{B_\nu\big[T_\mathrm{CMB}(z)\big] }{B_\nu\big[\tex\big] }\,.
\label{line_contrast_eq}
\end{align}
This is similar to eq.~\ref{continuum_contrast} and reflects the fact that we cannot detect the emission from lines where the excitation temperature is the same as the background radiation (as also included in, e.g.~\citealt{Scoville1974,vanderTak2007,Obreschkow2009})\footnote{We note that CO is unlikely to be observed in absorption against the CMB at any redshift, as the ubiquitous CMB radiation will effectively heat any gas (either via dust or line heating) to the temperature of the CMB.}.

The previous equations show that the observed intensity of a CO line depends on the optical depth $\tau_\nu^{J_u}$ and the excitation temperature \tex\ of each transition, which are set by the statistical equilibrium equation (eq.~\ref{stateq}) and the radiative transfer equation (eq.~\ref{rtline}). Computing $\tau_\nu^{J_u}$ and \tex\ is a non-trivial problem since the level populations and the radiation field are coupled. A commonly used method to deal with this is the `escape probability method' which gives the probability $\beta$ that a photon emitted in the transition escapes from the cloud. In this case, the radiation field becomes $\tilde U (1-\beta)$, and the way $\beta$ depends on the optical depth $\tau_\nu^{J_u}$ is specified by the geometry. The most widely used approximation is the large velocity gradient (LVG) approach, which assumes radially expanding spheres characterized by a velocity gradient $dv/dr$ along the line of sight (e.g.~\citealt{Scoville1974,Weiss2005,vanderTak2007}).

\subsection{The LTE case}
\label{co_lte}

If collisions dominate, i.e. $C_{ul} \gg A_{ul}$ (this is the case for high densities), then $\tex=T_\mathrm{kin}$ for all transitions. This is the local thermal equilibrium (LTE) case, and it implies that all the lines are thermalized (e.g.~\citealt{Obreschkow2009}).
As mentioned above, we assume thermodynamic equilibrium between the gas and the dust, i.e. $\tkin=\tdust$. Since in LTE, $\tex=\tkin$, then $\tex=\tdust$, and therefore the excitation temperature increases with redshift in the same way as the dust temperature (eq.~\ref{tdustz}).

\begin{figure}
\begin{minipage}{\linewidth}
\centering
\includegraphics[width=0.975\textwidth]{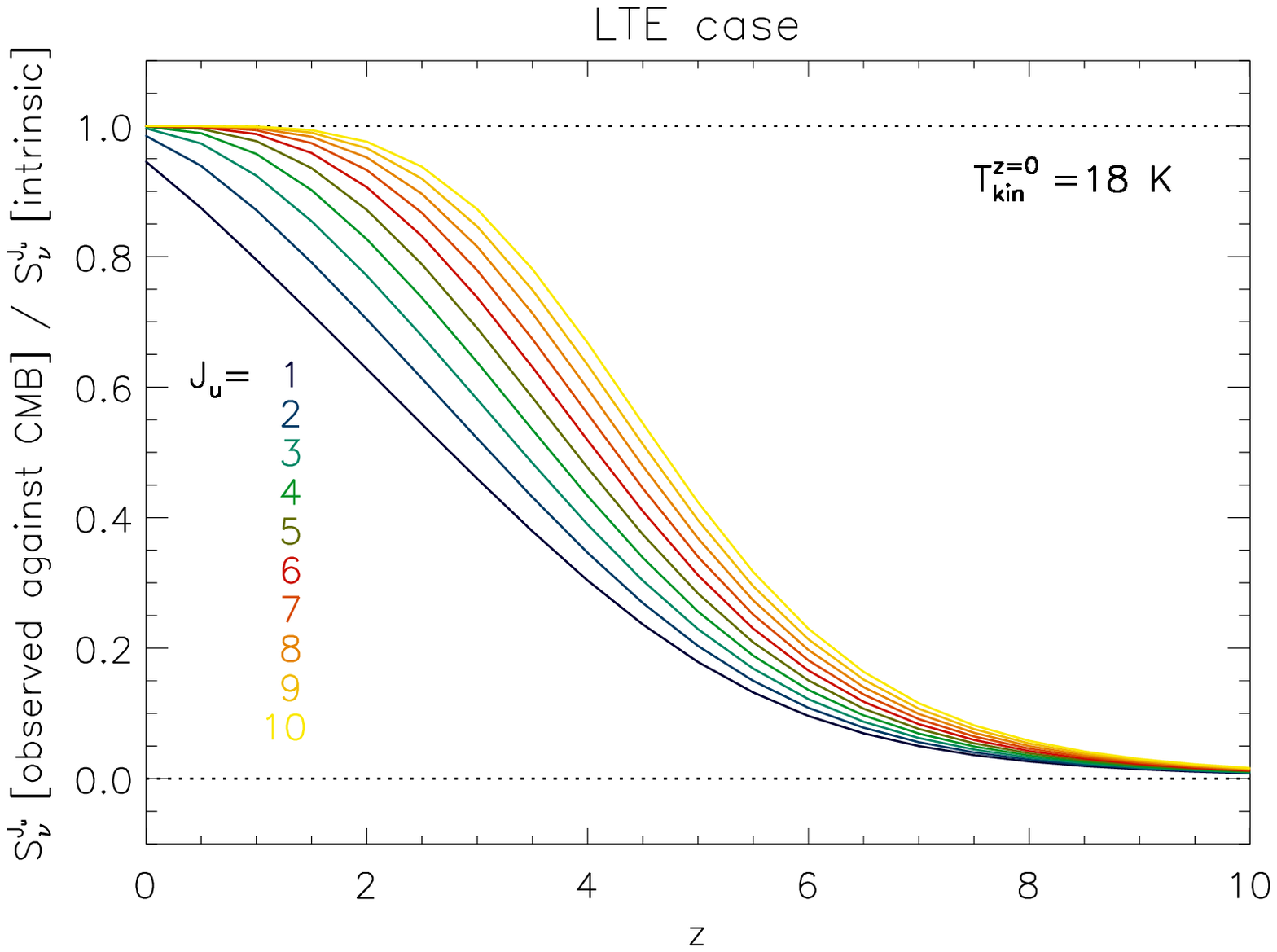}
\end{minipage}
\begin{minipage}{\linewidth}
\centering
\includegraphics[width=0.975\textwidth]{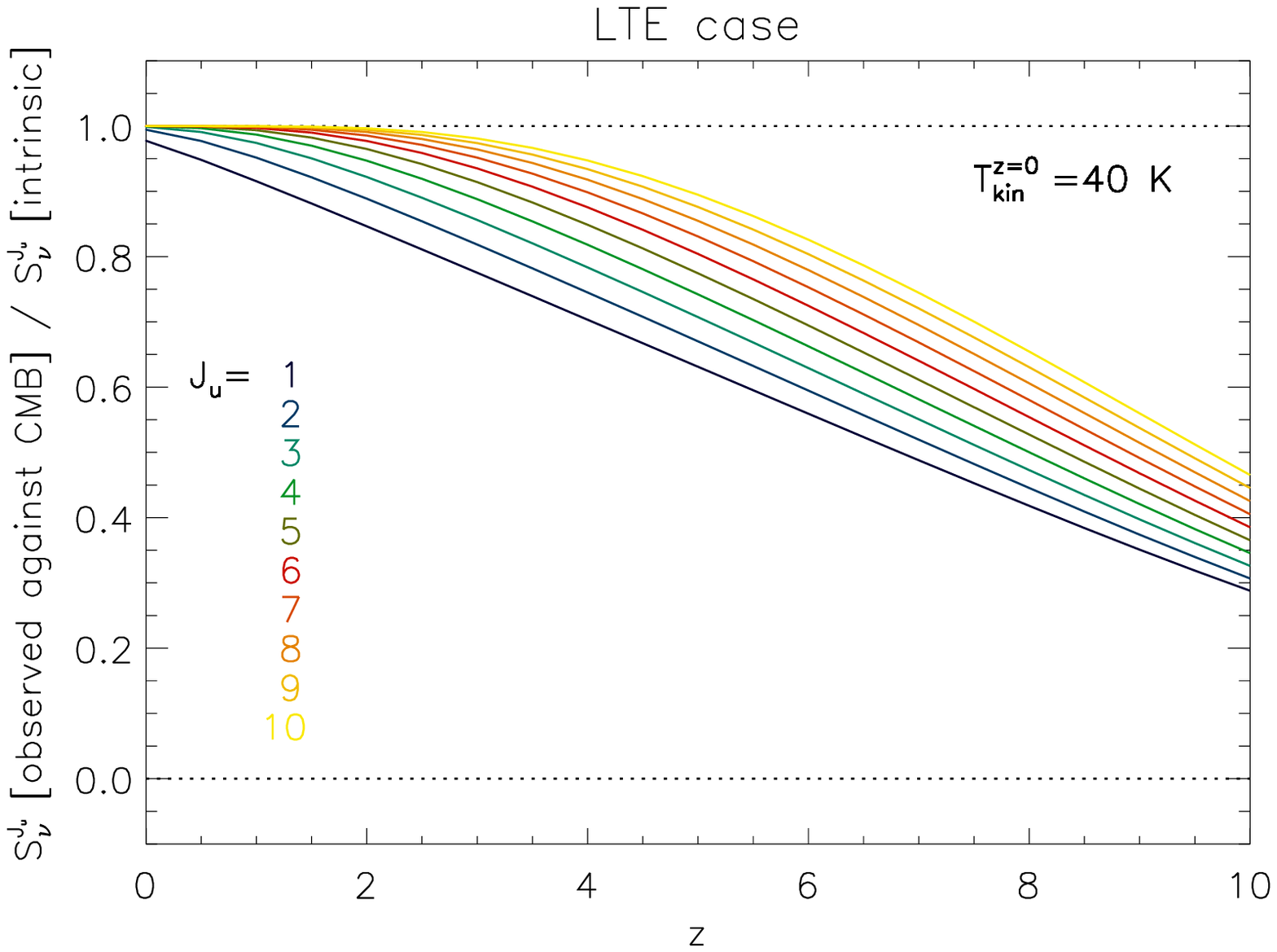}
\end{minipage}
\caption{Ratio between the line (velocity-integrated) flux observed against the CMB background and the intrinsic line flux for different CO transitions from $J_u=1$ to $J_u=10$ (shown with different colors). In the top panel, the intrinsic gas kinematic temperature is $\tkino=18$~K; in the bottom panel, we show a case with higher intrinsic gas temperature, $\tkino=40$~K. Since this is the LTE case, the excitation temperature \tex\ is the same for all levels and equal to $T_\mathrm{kin}$.
These plots include both effects of increasing CMB temperature with redshift on observed CO lines: (i) the increase of gas kinematic temperature (and hence the excitation temperature) as described by eq.~\protect\ref{tdustz}; (ii) the growing importance of the CMB as an observing background with redshift. When $S_\nu^{J_u}\mathrm{[observed~against~the~CMB]}/S_\nu^{J_u}\mathrm{[intrinsic]}=1$, the measured line flux is the total intrinsic flux emitted by the source; when $S_\nu^{J_u}\mathrm{[observed~against~the~CMB]}/S_\nu^{J_u}\mathrm{[intrinsic]}=0$, the line cannot be distinguished from the CMB background.}
\vspace{0.5cm}
\label{line_contrast}
\end{figure}

In Fig.~\ref{line_contrast}, we plot the ratio between the line luminosity observed against the CMB and the intrinsic luminosity for the CO transitions from $J_u=1$ to $J_u=10$ using eq.~\ref{line_contrast_eq}. We adopt two different intrinsic gas kinetic temperatures, $\tkino=18$~K (top panel) and $\tkino=40$~K (bottom panel). This figure shows that the decrease of CO line flux due to the CMB background can be a very significant effect, substantially decreasing the measured line flux for sources at high redshifts (see also \citealt{Combes1999,Obreschkow2009}). The effect of the CMB background is more pronounced for colder intrinsic excitation temperatures (which are closer to the CMB temperature at any redshift). This has strong implications for the detectability of CO lines from `quiescent' galaxies at high redshift, where the radiation field from star formation in the galaxy may lead to relatively low \tkin\ (and therefore \tex). For $\tkino=18$~K, typically less than half of the total line luminosity can be measured at $z>4$, with even lower fractions for the lowest-$J_u$ transitions. For higher gas excitation temperatures, the effect is not as strong. For $\tkino=40$~K, over 70\% of the intrinsic flux is recovered. The decrease of measured CO line flux at high redshift shown in these plots has strong implications for measuring the molecular gas ($M_\mathrm{H_2}$) mass of high-redshift galaxies using the CO line luminosity, $L_\mathrm{CO}$. Since $M_\mathrm{H_2} \propto L_\mathrm{CO}$ (e.g.~\citealt{Solomon2005}), if we do not correct for the CMB in order to obtain the intrinsic CO luminosity (using eq.~\ref{line_contrast_eq}), then we may severely underestimate $L_\mathrm{CO}$ and consequently underestimate the molecular gas mass of the galaxies (by more than a factor of 2 at $z>4$ for 18~K gas).

\begin{figure}
\begin{center}
\includegraphics[width=0.5\textwidth]{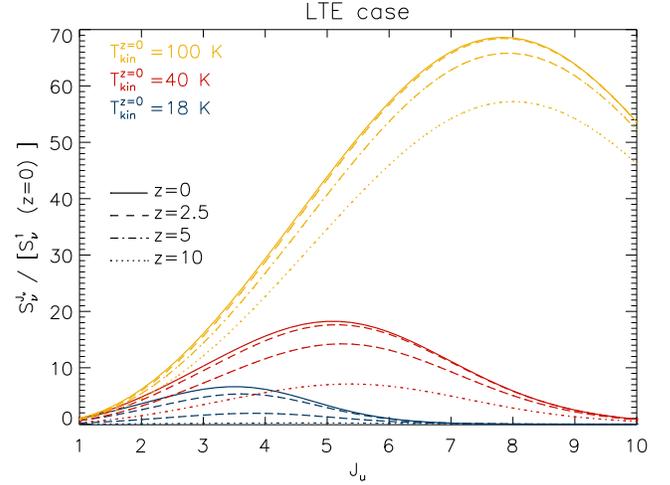}
\caption{Variation of observed CO SLEDs with redshift for three different intrinsic kinetic temperatures: $\tkino=18$~K (blue), $40$~K (red), and $100$~K (yellow). For each intrinsic temperature, we consider four different redshifts of the emitting galaxy: $z=0$ (solid lines); $z=2$ (dashed lines); $z=5$ (dot-dashed lines); and $z=10$ (dotted lines). For each temperature, all the CO SLEDs are normalized to the flux of the CO(1--0) line at $z=0$, $S_\nu^1 (z=0)$.}
\label{co_sled}
\end{center}
\end{figure}

Fig.~\ref{line_contrast} also shows that the fraction of CO line luminosity that is measured against the CMB for a given gas excitation temperature is not the same for all CO lines as a function of redshift. This implies that the shape of the observed CO spectral line energy distributions (SLEDs) will change with redshift, even if the intrinsic shape, i.e. emitted SLED (which depends only on the gas properties and excitation) does not change. Like in the case of the dust SEDs, this can affect how one interprets the properties of high-redshift galaxies based on their observed CO SLEDs, namely the excitation conditions of the gas. We investigate how large this effect is by analyzing the CO SLEDs is as a function of intrinsic excitation temperature and redshift.
We consider three intrinsic kinetic temperatures, $\tkino=\tdusto=18$, $40$, and $100$~K. For each temperature, we build the expected CO SLEDs by computing the velocity-integrated flux of each $J_u>1$ transition normalized to the velocity-integrated flux of the $J_u=1$ transition at $z=0$ (this normalization highlights the fact that the lines become less bright due to lack of contrast with the CMB as shown in Fig.~\ref{line_contrast}). We use the first term of eq.~\ref{inu_line} to compute the intrinsic CO line luminosity of each transition, with the excitation temperature $\tex=\tkin$, and the line optical depth as given by equation 5 in \cite{Obreschkow2009} (derived for the LTE approximation):
\begin{equation}
\tau_\nu^{{J_u}}\mathrm{[LTE]}=7.2\,\tau_c \exp\bigg(-\frac{h\,\nuco\,J_u^2}{2\,k\,\tex}\bigg)\sinh\bigg(\frac{h\,\nuco\,J_u}{2\,k\,\tex}\bigg)\,,
\label{tau_line}
\end{equation}
where $\tau_c$ is a constant. Following \cite{Obreschkow2009}, we fix $\tau_c=2$ (which they show provides a good fit to observed CO SLEDs). We compute the expected CO line SLEDs for three intrinsic kinetic temperatures at different redshifts by including both the extra heating by the CMB that increases \tdust\ and hence \tkin\ and \tex, and the brighter CMB background as discussed above.
The results are plotted in Fig.~\ref{co_sled}, which shows two main effects: (i) the overall flux of the lines decreases at higher redshifts, as discussed above; (ii) the decrease in line intensity is not the same for all transitions, which slightly changes the shape of the SLEDs.
At fixed $\tkino$, there is a slight shift of the CO SLED peak towards higher $J_u$ with increased redshift. This is a consequence both the slightly higher excitation temperature (due to additional CMB heating) and the fact that the CMB background affects the low-$J_u$ line measurements more than the high-$J_u$ lines (see Fig.~\ref{line_contrast}). As expected, these effects are stronger for the lowest \tkino; gas with $\tkino=18$~K is practically undetectable at $z\gtrsim5$.

\subsection{Non-LTE examples}
\label{co_nonlte}

\begin{figure*}
\vspace{0.5cm}
\begin{minipage}{0.5\linewidth}
\centering
\includegraphics[width=0.95\textwidth]{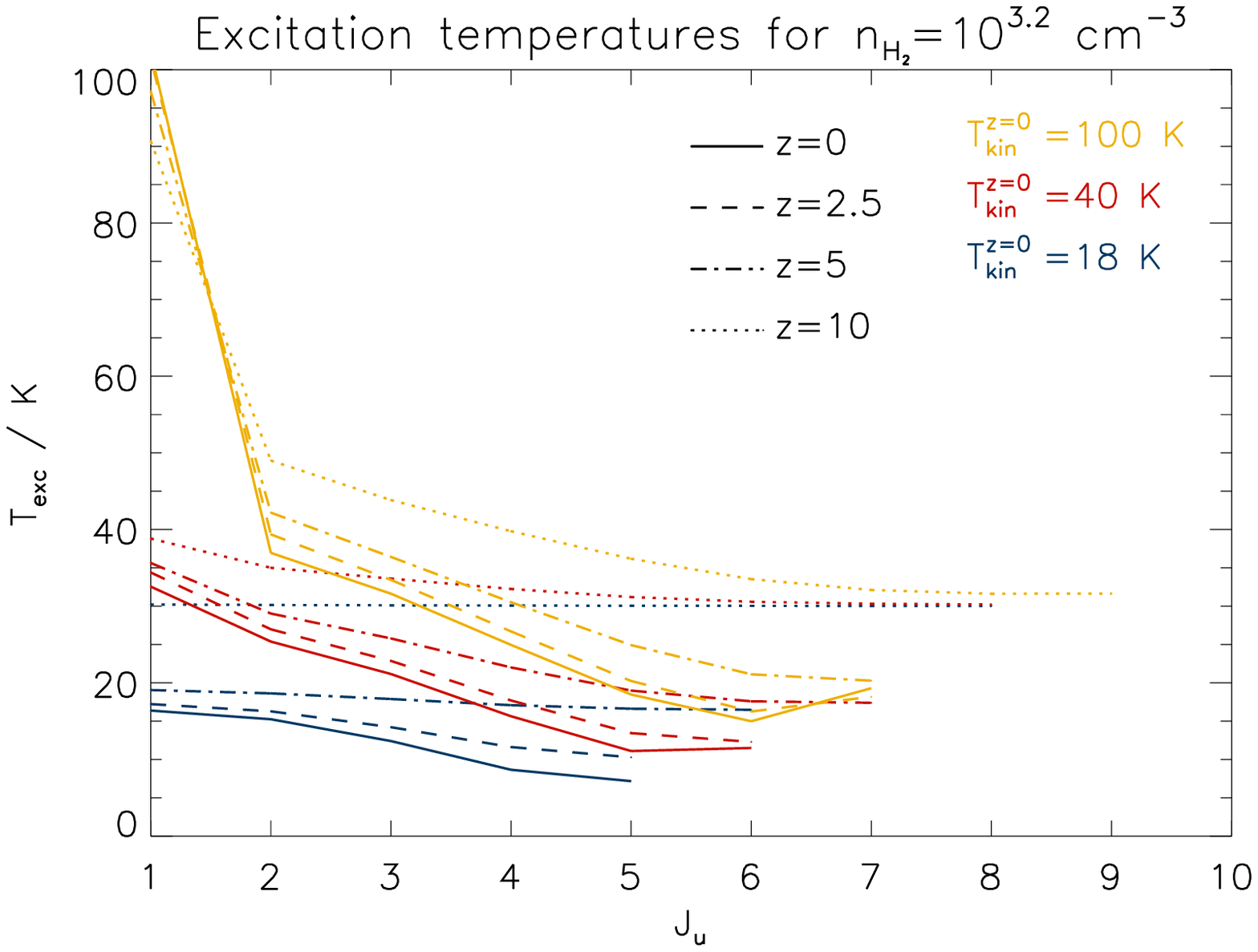}
\end{minipage}
\begin{minipage}{0.5\linewidth}
\centering
\includegraphics[width=0.95\textwidth]{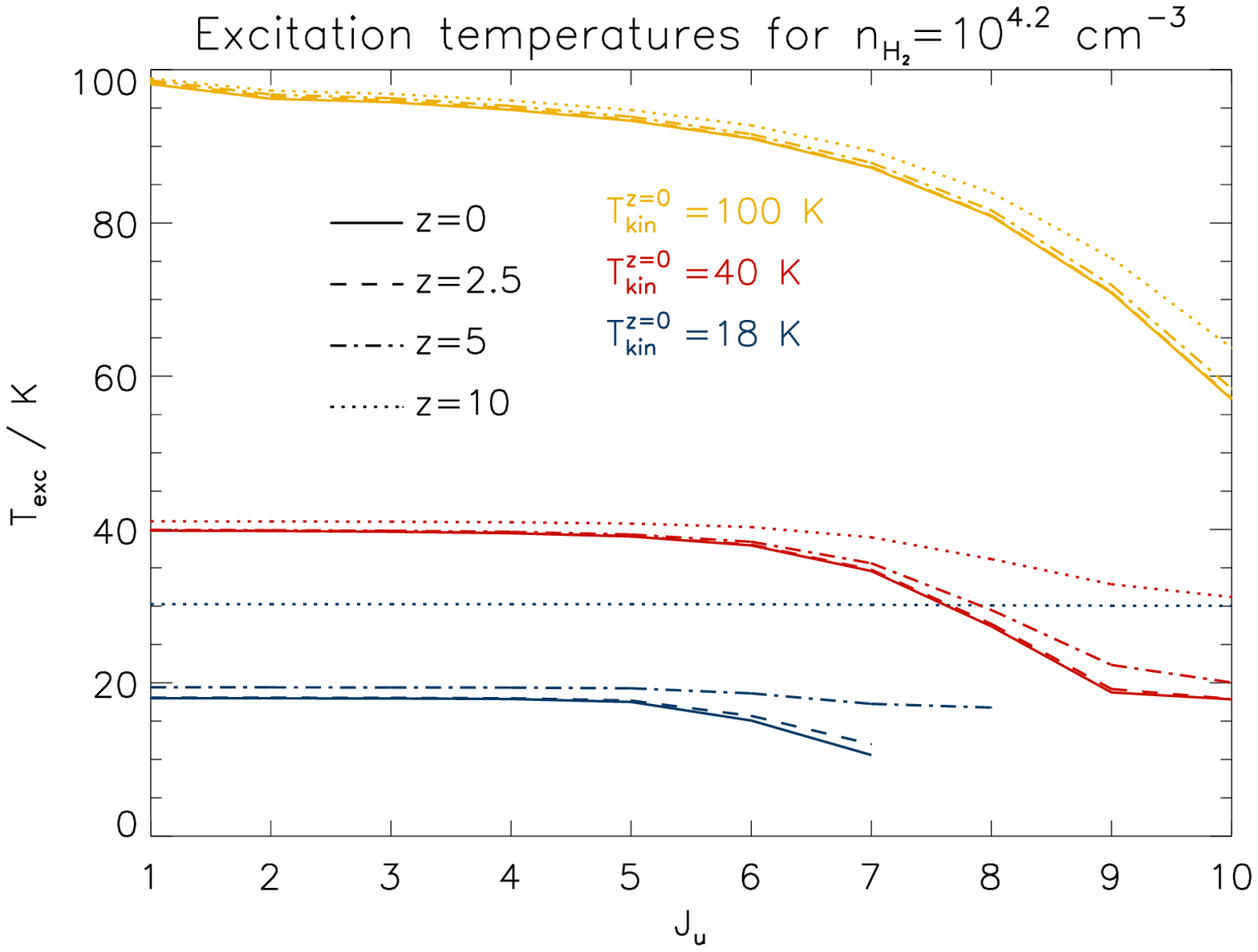}
\end{minipage}
\caption{Excitation temperatures of the various CO transitions computed using the LVG approximation (using the models of \citealt{Weiss2005}), for a low-density model (left-hand panel) and a high-density model (right-hand panel). We consider three intrinsic kinetic temperatures at $z=0$: $\tkino=18$ (blue), $40$ (red), and $100$~K (yellow). For each model, the kinetic temperature increases with redshift following the dust temperature (eq.~\ref{tdustz}, which includes extra dust heating by the CMB at high redshift). The resulting excitation temperatures at different redshifts are plotted with different line styles: $z=0$ (solid), $z=2.5$ (dashed), $z=5$ (dot-dashed), and $z=10$ (dotted). We plot only transitions where the optical depth  $\tau_\nu^{J_u}$ is higher than 0.1 (Fig.~\ref{tau_lvg}).}
\vspace{0.5cm}
\label{texc_lvg}
\end{figure*}

\begin{figure*}
\begin{minipage}{0.5\linewidth}
\centering
\includegraphics[width=0.95\textwidth]{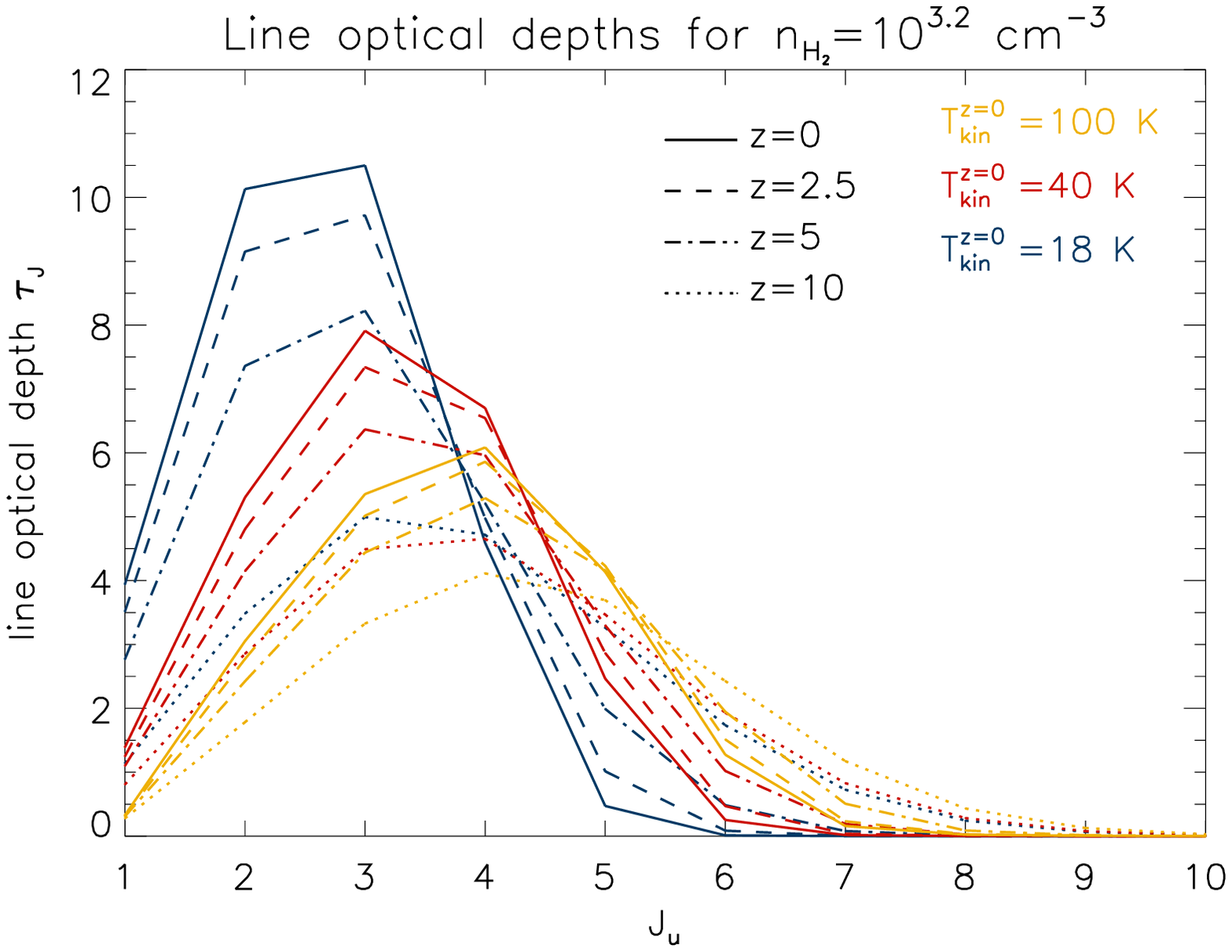}
\end{minipage}
\begin{minipage}{0.5\linewidth}
\centering
\includegraphics[width=0.95\textwidth]{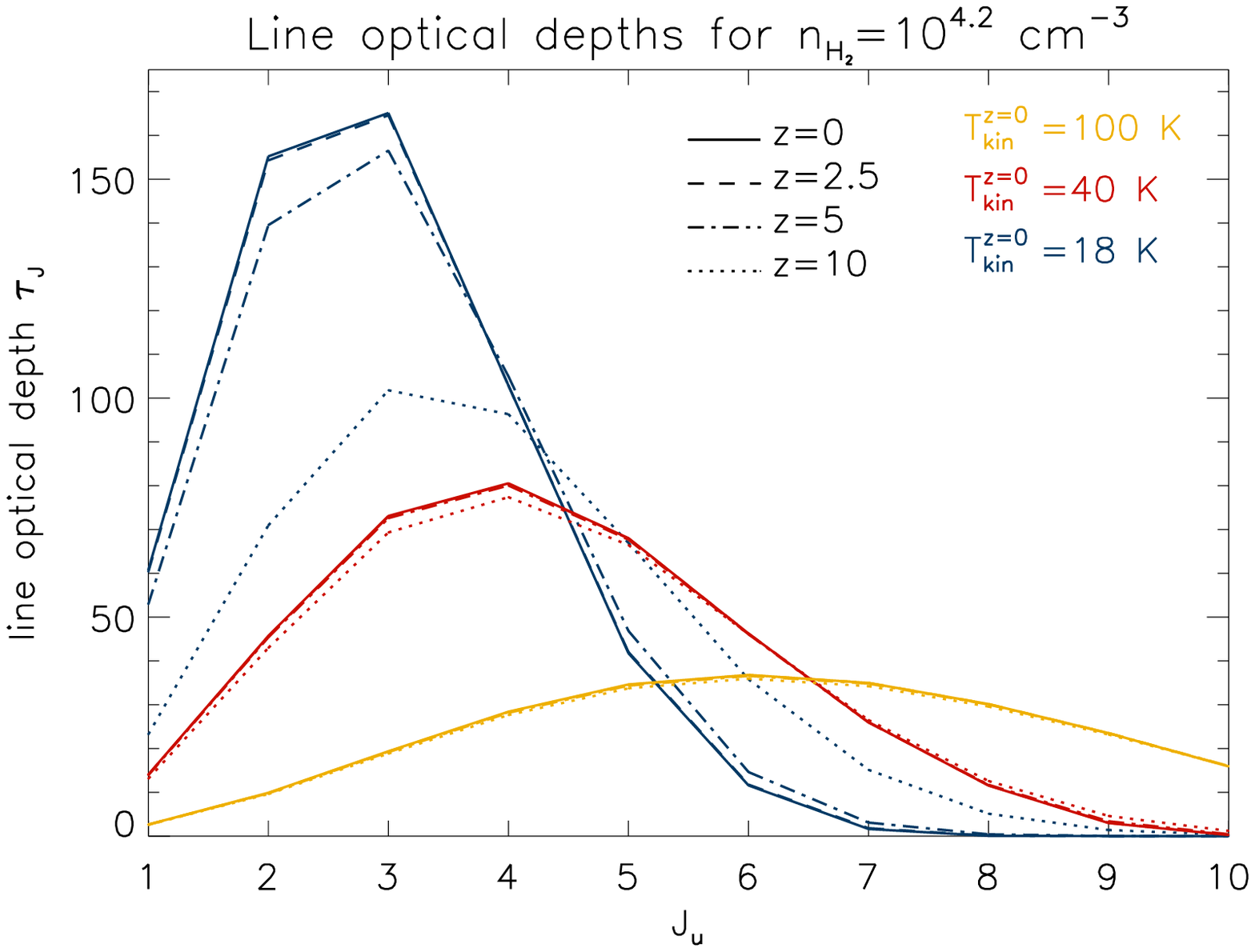}
\end{minipage}
\caption{Optical depths of the various CO transitions computed using the LVG approximation \citep{Weiss2005}, for a low-density model (left-hand panel) and a high-density model (right-hand panel). The color and line meanings are the same as in Fig.~\ref{texc_lvg}.}
\vspace{0.5cm}
\label{tau_lvg}
\end{figure*}

\begin{figure*}
\begin{minipage}{0.5\linewidth}
\centering
\includegraphics[width=0.95\textwidth]{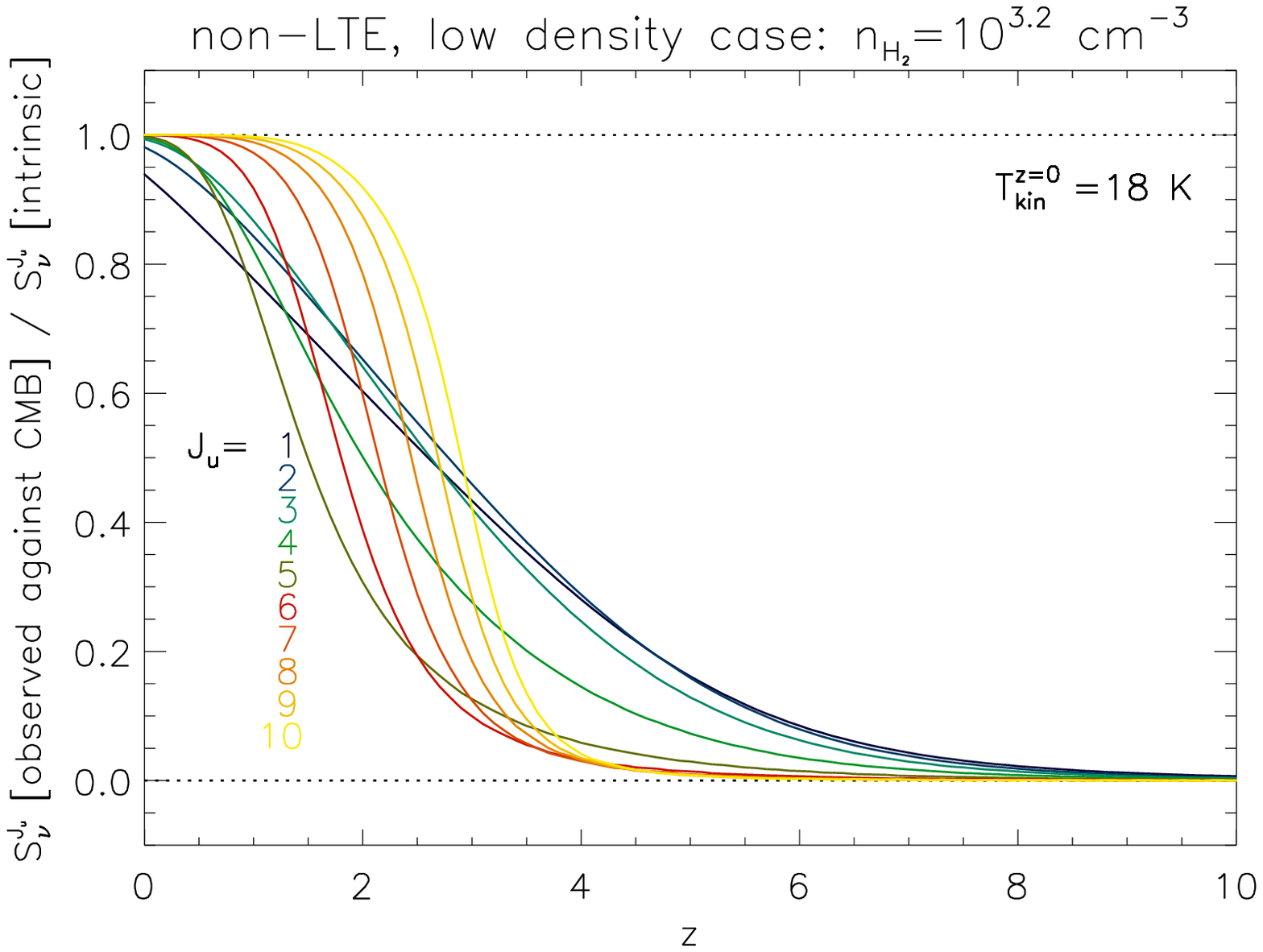}
\end{minipage}
\begin{minipage}{0.5\linewidth}
\centering
\includegraphics[width=0.95\textwidth]{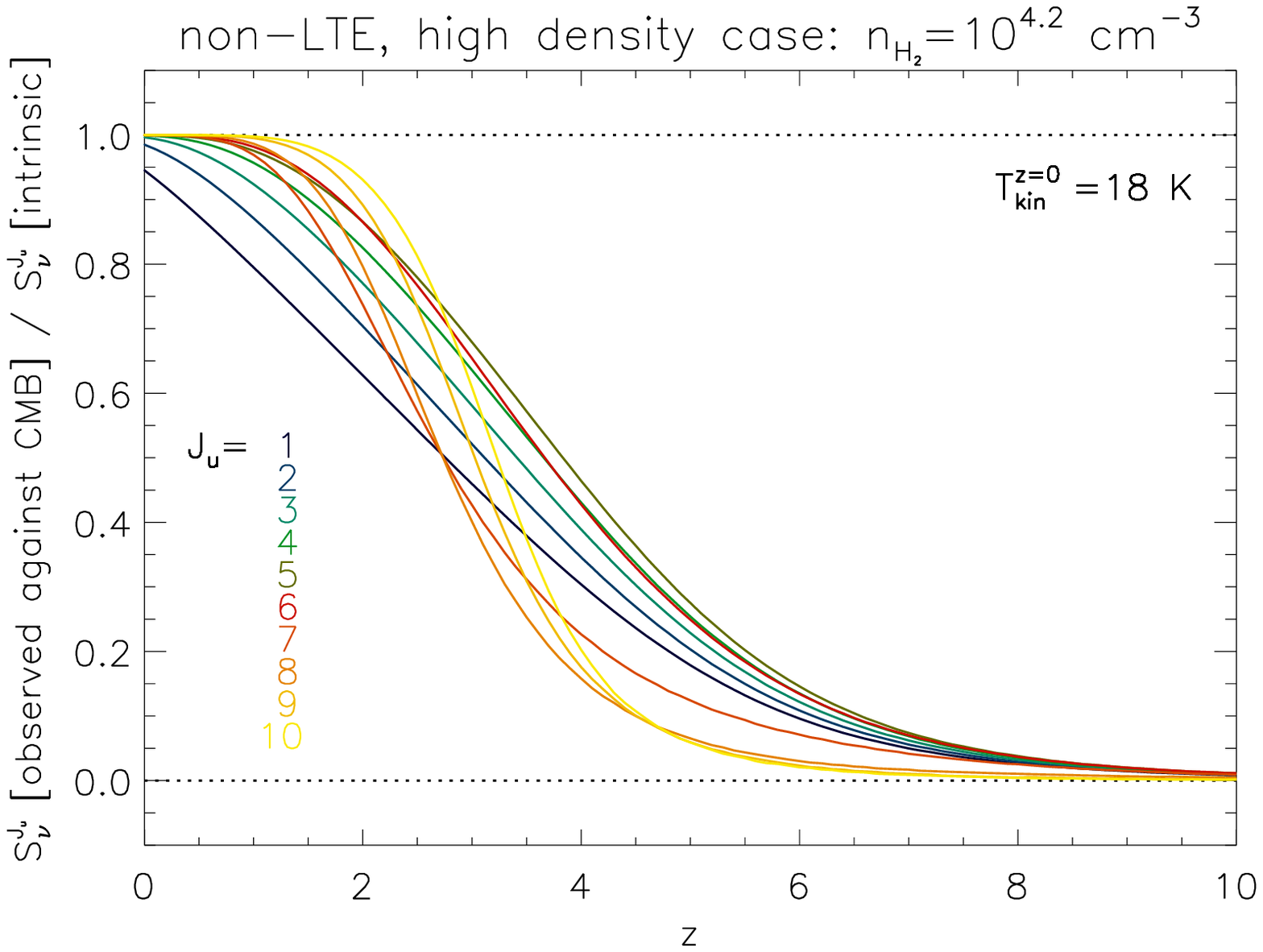}
\end{minipage}
\begin{minipage}{0.5\linewidth}
\centering
\includegraphics[width=0.95\textwidth]{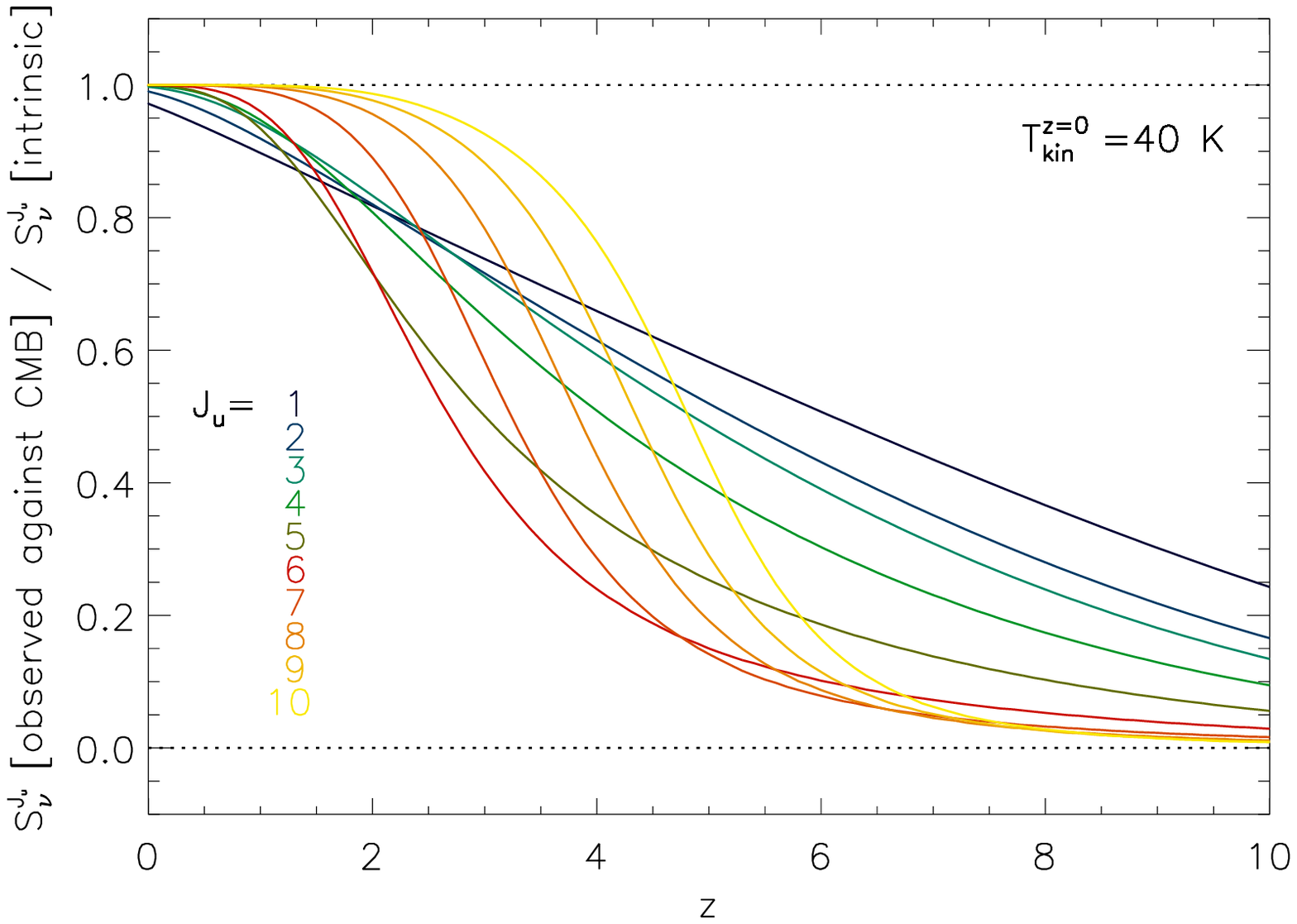}
\end{minipage}
\begin{minipage}{0.5\linewidth}
\centering
\includegraphics[width=0.95\textwidth]{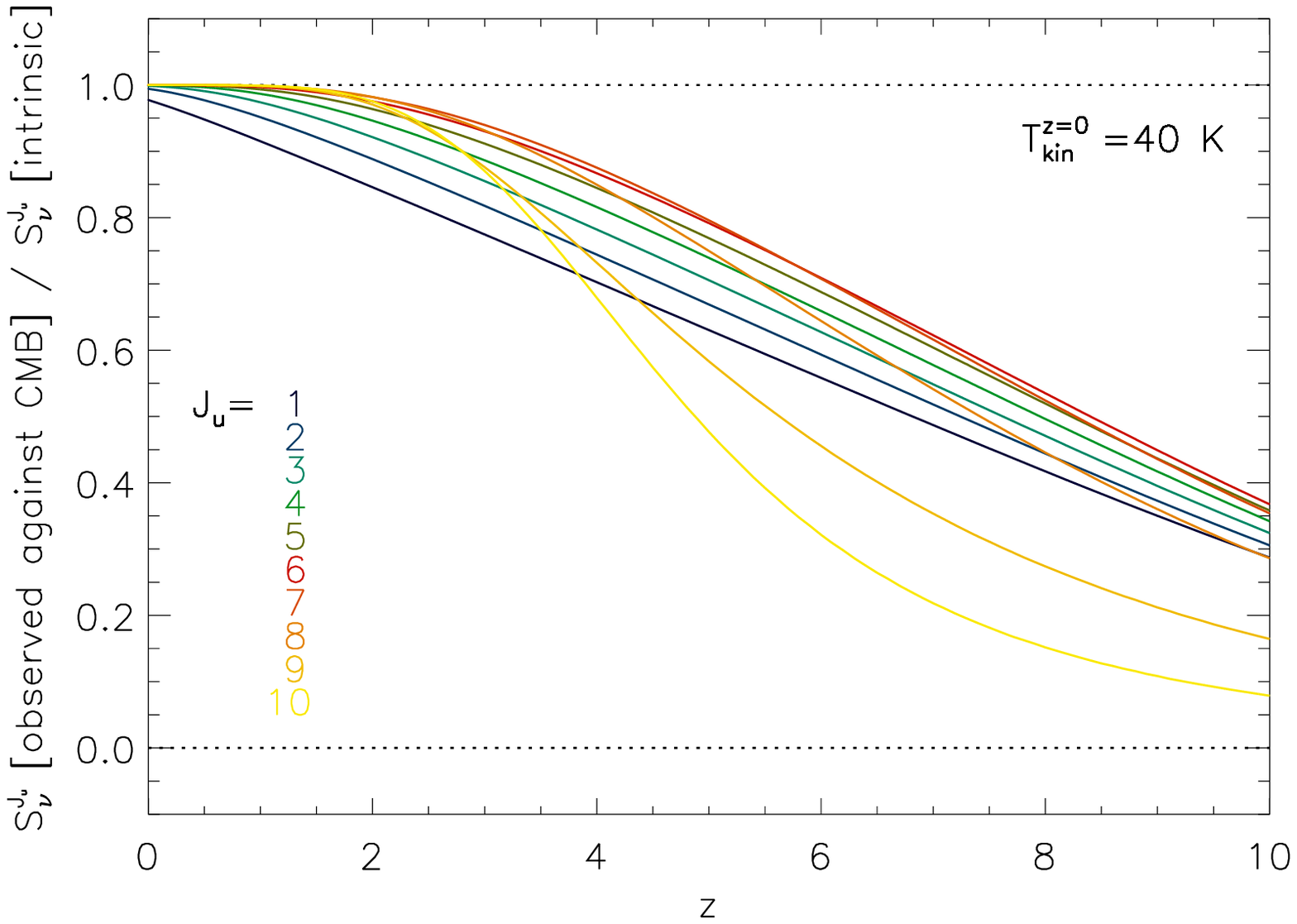}
\end{minipage}
\caption{Ratio between the line luminosity observed against the CMB background and the intrinsic line luminosity for different CO transitions from $J_u$ in the non-LTE examples described in Section~\ref{co_nonlte} (shown with different colors as in Fig~\ref{line_contrast}). In the top panels we show models with intrinsic gas kinematic temperature $\tkino=18$~K; in the bottom panel, we show the cases with higher intrinsic gas temperature, $\tkino=40$~K. The excitation temperature for each transition is computed using the \cite{Weiss2005} LVG model, with the parameters described in Section~\ref{co_nonlte}: the left-hand panels correspond to the low-density case, and the right-hand panels correspond to the high-density case. }
\vspace{0.5cm}
\label{line_contrast_lvg}
\end{figure*}

\begin{figure*}
\vspace{0.5cm}
\begin{minipage}{0.5\linewidth}
\centering
\includegraphics[width=0.95\textwidth]{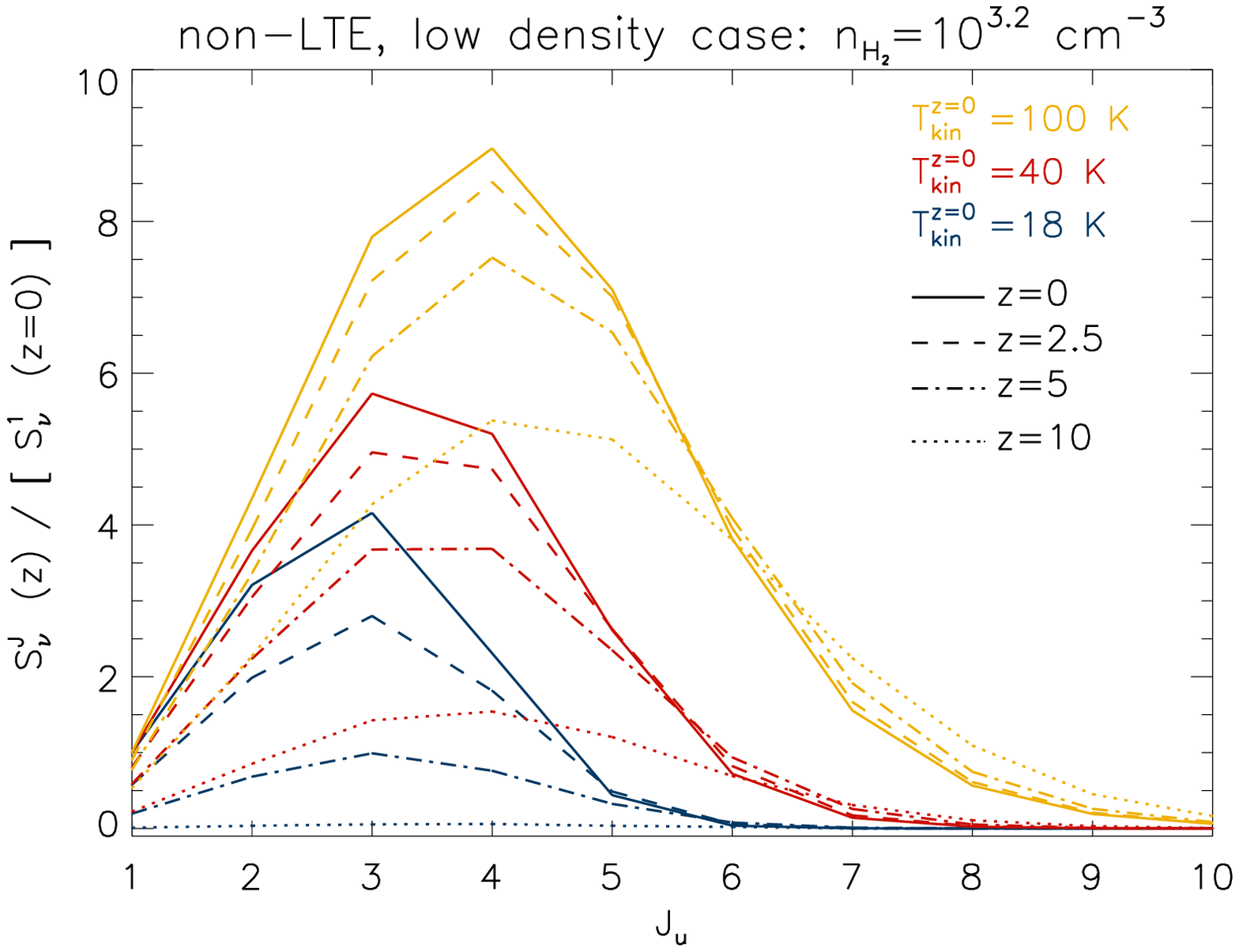}
\end{minipage}
\begin{minipage}{0.5\linewidth}
\centering
\includegraphics[width=0.95\textwidth]{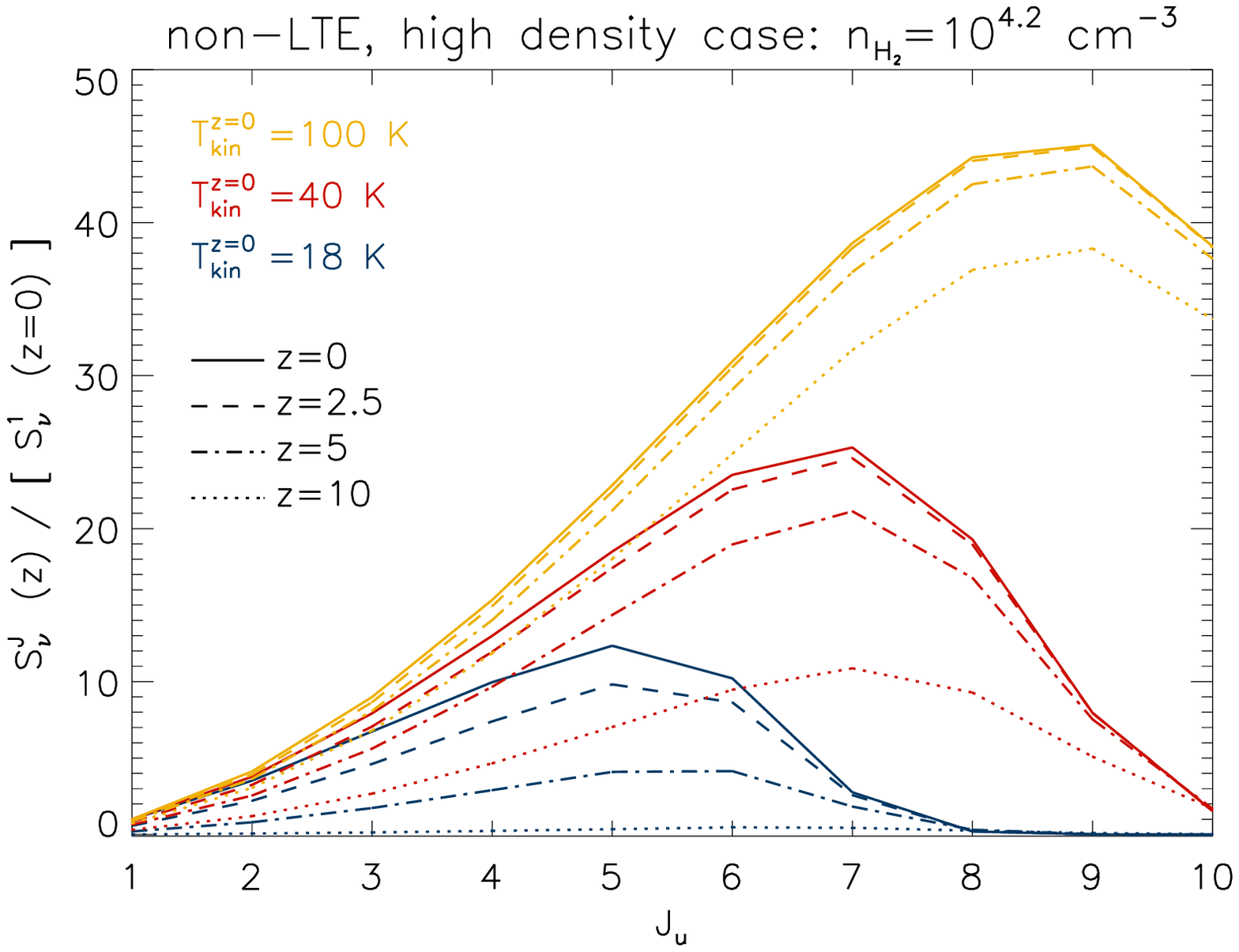}
\end{minipage}
\caption{Variation of observed CO SLEDs with redshift for three different intrinsic kinetic temperatures: $\tkino=18$~K (blue), $40$~K (red), and $100$~K (yellow), computed using the \cite{Weiss2005} LVG model for the low-density case (left-hand panel) and the high-density case (right-hand case). For each intrinsic temperature, we consider four different redshifts of the emitting galaxy: $z=0$ (solid lines); $z=2$ (dashed lines); $z=5$ (dot-dashed lines); and $z=10$ (dotted lines). For each temperature, all the CO SLEDs are normalized to the flux of the CO(1--0) line at $z=0$, $S_\nu^1 (z=0)$.}
\vspace{0.5cm}
\label{co_sled_lvg}
\end{figure*}

In real galaxies, the molecular gas is not likely to be in LTE, and so the radiative transfer of the lines and the population levels have to be solved simultaneously as described in Section~\ref{co_general}. In this section, we show how the CMB affects the detectability of the CO lines, where the line fluxes and optical depths are computed using the LVG model of \cite{Weiss2005}, which assumes an expanding sphere geometry with velocity gradient $dv/dr$. 
The main free parameters of this model are:
\begin{itemize}
\item The kinetic temperature of the gas, \tkin. As in the previous section, we assume $\tkin(z)=\tdust(z)$, i.e. that there is a 100\% effective coupling of the dust and gas temperatures. We analyze three values of $\tkino$: 18, 40 and 100~K.
\item The number density of H$_2$ molecules, $n_\mathrm{H_2}$. We analyze two examples:
\begin{itemize}
\item `low-density' case: $n_\mathrm{H_2}=10^{3.2}$~cm$^{-3}$
\item `high-density' case: $n_\mathrm{H_2}=10^{4.2}$~cm$^{-3}$
\end{itemize}
\item The number density of CO molecules. We assume $n_\mathrm{CO}=8\times10^{-5}~n_\mathrm{H_2}$.
\item The velocity gradient $dv/dr$. Assuming that the clouds are virialized, we obtain $dv/dr=8$~km~s$^{-1}$~pc$^{-1}$ for the low density case and  $dv/dr=3.9$~km~s$^{-1}$~pc$^{-1}$ for the high density case.
\end{itemize}
Using these parameters, the excitation temperature and optical depth of each line are computed by coupling the statistical equilibrium and radiative transfer equations (Section~\ref{co_general}; \citealt{Weiss2005}). 
In Figs.~\ref{texc_lvg} and \ref{tau_lvg}, we plot the variation of excitation temperatures \tex\ and line optical depths $\tau_\nu^{J_u}$, respectively, with redshift and input kinetic temperature \tkino, resulting from these calculations.
These plots show that the high-density case approaches LTE for low $J_u$: $\tex \simeq \tkin$ up to $J_u \simeq 5$ i.e. the low-$J_u$ lines are thermalized (i.e. the H$_2$ density is higher than the critical density for these levels). For higher $J_u$, $\tex < \tkin$ because at these densities and kinetic temperatures these levels are not as populated (as they would be in the LTE case).
We also note that, for the lowest $\tkino$, at high redshift, the kinetic temperature becomes very close to the CMB temperature. Due to the strength of the CMB at high redshifts, the CMB defines the minimum excitation temperature as the CO levels become radiatively dominated. In this case, the radiative processes dominate and so the excitation temperature gets close to $T_\mathrm{rad}$ i.e. \tcmb.
Finally, these plots also show that $\tex < \tkin$ at low density and high $J_u$, i.e. less molecules are in excited states because there is less collisional excitation. At higher redshifts, \tex\ is higher because \tkin\ is larger due to the dust being hotter (thanks to extra CMB heating) and, at the same time, CMB radiative excitation becoming more important.

Once the excitation temperature and optical depth of each transition are determined using the LVG models, we can compute the velocity-averaged flux of each line as described in the previous sections.
In Fig.~\ref{line_contrast_lvg}, we show the ratio between the line luminosity observed against the CMB background and the intrinsic line luminosity for different CO transitions for the low- and high-density case, for intrinsic gas kinetic temperatures \tkino\ of 18 and 40~K. These plots can be directly compared with the LTE case shown in Fig.~\ref{line_contrast}. As for the LTE case, the contrast between the lines and the CMB decreases with redshift, and it decreases more rapidly for the low-temperature case, since the temperature is closer to the CMB temperature. The main difference between these plots and the LTE case are the high-$J_u$ transitions. Since the higher-$J_u$ levels are underpopulated compared to the LTE case (the lines are not thermalized), their excitation temperature is lower than $\tkin(z)$ (see Fig.~\ref{texc_lvg}), and so the contrast between these lines and the CMB is even lower than for the low-$J_u$ lines (contrary to what happens in the LTE case). The predicted decrease in contrast is stronger for $\tkino=18$~K, specially for low H$_2$ density: in this case, lines with $J_u>5$ become practically undetectable against the CMB from $z=4$.
For the lowest-$J_u$ transitions, the contrast behaves similar to the LTE case, because the low-$J_u$ are thermalized.

In Fig.~\ref{co_sled_lvg}, we plot the CO SLEDs computed using the LVG code, i.e. with the excitation temperatures and line optical depths shown in Figs.~\ref{texc_lvg} and \ref{tau_lvg}. As for the LTE case (Fig.~\ref{co_sled}), these plots show how the flux of the lines decreases with redshift due to decreasing contrast with the CMB (and also to some extent different line optical depths; Fig.~\ref{tau_lvg}). In this case it is harder to distinguish differences between the CO SLED shapes at different redshifts caused by the difference in contrast against the CMB or by different intrinsic values of \tex\ and $\tau_\nu^{J_u}$ at different redshifts (set by increasing increasing excitation by the CMB both via extra dust heating and radiative processes; Figs.~\ref{texc_lvg} and \ref{tau_lvg}), which would produce different intrinsic CO SLEDs at different redshifts. However, we have shown that the effect of the CMB background is often non-negligible (Figs.~\ref{line_contrast_lvg}) and must be taken into account when interpreting observed SLEDs using LVG models (e.g.~\citealt{Scoville1974,Weiss2005,vanderTak2007}).

We note that if we do not assume a perfect coupling between the dust and gas temperatures, i.e. that \tkin($z$)=\tdust($z$), the effects presented in this section will be even more pronounced, i.e. in general the lines look weaker. Indeed, if we assume that $\tkin$ to be constant with redshift (i.e. $\tkin(z)=\tkino$), at the high redshifts we obtain $\tkin(z)<\tdust(z)$, since \tdust\ increases with $z$ due to the extra CMB heating (eq.~\ref{tdustz}). In this case, using our LVG modelling, we obtain the result that \tex\ tends to be lower in general (because since \tkin\ is smaller the collisions are not as effective bringing molecules to excited levels), and so the contrast between the lines and the CMB is weaker.

%***************************************************************************
\section{Summary \& Conclusion}
\label{conclusion}

In this paper, we have analyzed in detail how the CMB affects (sub-)millimeter observations of both the dust continuum and CO lines in high-redshift galaxies. We have shown that, at high redshifts, the CMB becomes an increasingly significant additional heating source for the dust and gas, boosting their temperatures and enhancing the dust continuum and CO line emission of the galaxies. However, we show that at higher redshifts, since the CMB is hotter and therefore brighter, the contrast of the intrinsic dust (and line) emission against the CMB decreases. We have quantified how this affects the detectability of dust and lines in specific cases with fixed temperatures. Additionally, we provide general correction factors to compute what fraction of the intrinsic dust (and line) emission can be detected against the CMB as a function of frequency, redshift and temperature, and a recipe to include these effects when comparing dust emission models with observations.

We show that neglecting the CMB effect on (sub-)mm observations of high-redshift galaxies can lead to significant differences in the derived intrinsic properties of dust, in particular a severe underestimation of the dust mass. Specifically, the intrinsic dust mass of cool dust ($\tdust\lesssim 20$~K) at $z>5$ can be underestimated by at least one order of magnitude. The inferred total dust luminosity (and hence star formation rate) is less drastically affected: between 30\% and 50\%.
Similarly, if the effect of the CMB as an observing background is neglected when measuring CO lines, this can lead to wrong interpretations of the molecular gas properties (such as total mass, density and kinetic temperature) via the observed CO luminosity and the SLEDs. For low gas kinetic temperatures (18~K) and different density scenarios, we find that, at $z>5$, less than 20\% of the intrinsic CO(1--0) or CO(2--1) line fluxes can be measured against the CMB. This implies that, without proper corrections, the inferred molecular gas mass would be 20\% of the intrinsic value. At higher gas kinetic temperature (40~K), the underestimation of the gas mass would be less dramatic since between 20\% and 60\% of the intrinsic line fluxes are measured (depending on the density), however this is still significant when studying the gas reservoir/star formation efficiency of high-redshift galaxies.

Finally, our results imply that the cold ISM of galaxies at high redshift, with intrinsic temperatures of about 20~K, if existent, will be difficult to measure even with the unprecedented sensitivity of modern (sub-)millimeter observatories such as ALMA, simply because the contrast between the continuum (and line) emission against the CMB background decreases dramatically at $z\gtrsim4$.

%******************************************************************************

\section*{Acknowledgements}

We thank the referee for a careful reading of the manuscript. We are also grateful to Fran\c{c}oise Combes, Nick Scoville and Alberto Bolatto for comments on the manuscript.
EdC acknowledges funding through the ERC grant `Cosmic Dawn'.
RD acknowledges funding from Germany's national research center for aeronautics and space (DLR, project FKZ 50 OR 1104).

%******************************************************************************

% Bibliography and bibfile
\def\aj{AJ}
\def\araa{ARA\&A}
\def\apj{ApJ}
\def\apjl{ApJ}
\def\apjs{ApJS}
\def\apss{Ap\&SS}
\def\aap{A\&A}
\def\aapr{A\&A~Rev.}
\def\aaps{A\&AS}
\def\mnras{MNRAS}
\def\pasp{PASP}
\def\pasj{PASJ}
\def\qjras{QJRAS}
\def\nat{Nature}

\def\aplett{Astrophys.~Lett.}
\def\aas{AAS}
\let\astap=\aap
\let\apjlett=\apjl
\let\apjsupp=\apjs
\let\applopt=\ao

\bibliographystyle{apj}
%\bibliography{bib_dacunha}

\begin{thebibliography}{26}
\expandafter\ifx\csname natexlab\endcsname\relax\def\natexlab#1{#1}\fi

\bibitem[{{Blain}(1999)}]{Blain1999}
{Blain}, A.~W. 1999, \mnras, 309, 955

\bibitem[{{Blain} {et~al.}(2002){Blain}, {Smail}, {Ivison}, {Kneib}, \&
  {Frayer}}]{Blain2002}
{Blain}, A.~W., {Smail}, I., {Ivison}, R.~J., {Kneib}, J.-P., \& {Frayer},
  D.~T. 2002, \physrep, 369, 111

\bibitem[{{Combes} {et~al.}(1999){Combes}, {Maoli}, \& {Omont}}]{Combes1999}
{Combes}, F., {Maoli}, R., \& {Omont}, A. 1999, \aap, 345, 369

\bibitem[{{da Cunha} {et~al.}(2008){da Cunha}, {Charlot}, \&
  {Elbaz}}]{daCunha2008}
{da Cunha}, E., {Charlot}, S., \& {Elbaz}, D. 2008, \mnras, 388, 1595

\bibitem[{{Daddi} {et~al.}(2010){Daddi}, {Elbaz}, {Walter}, {Bournaud},
  {Salmi}, {Carilli}, {Dannerbauer}, {Dickinson}, {Monaco}, \&
  {Riechers}}]{Daddi2010b}
{Daddi}, E., {Elbaz}, D., {Walter}, F., {et~al.} 2010, \apjl, 714, L118

\bibitem[{{Draine}(2011)}]{Draine2011}
{Draine}, B.~T. 2011, {Physics of the Interstellar and Intergalactic Medium}

\bibitem[{{Draine} \& {Lee}(1984)}]{Draine1984}
{Draine}, B.~T., \& {Lee}, H.~M. 1984, \apj, 285, 89

\bibitem[{{Draine} \& {Li}(2007)}]{Draine2007}
{Draine}, B.~T., \& {Li}, A. 2007, \apj, 657, 810

\bibitem[{{Dunne} {et~al.}(2000){Dunne}, {Eales}, {Edmunds}, {Ivison},
  {Alexander}, \& {Clements}}]{Dunne2000}
{Dunne}, L., {Eales}, S., {Edmunds}, M., {et~al.} 2000, \mnras, 315, 115

\bibitem[{{Genzel} {et~al.}(2010){Genzel}, {Tacconi}, {Gracia-Carpio},
  {Sternberg}, {Cooper}, {Shapiro}, {Bolatto}, {Bouch{\'e}}, {Bournaud},
  {Burkert}, {Combes}, {Comerford}, {Cox}, {Davis}, {Schreiber},
  {Garcia-Burillo}, {Lutz}, {Naab}, {Neri}, {Omont}, {Shapley}, \&
  {Weiner}}]{Genzel2010}
{Genzel}, R., {Tacconi}, L.~J., {Gracia-Carpio}, J., {et~al.} 2010, \mnras,
  407, 2091

\bibitem[{{Groves} {et~al.}(2012){Groves}, {Krause}, {Sandstrom}, {Schmiedeke},
  {Leroy}, {Linz}, {Kapala}, {Rix}, {Schinnerer}, {Tabatabaei}, {Walter}, \&
  {da Cunha}}]{Groves2012}
{Groves}, B., {Krause}, O., {Sandstrom}, K., {et~al.} 2012, \mnras, 426, 892
  
\bibitem[{{Klaas} {et~al.}(2001){Klaas}, {Haas}, {M{\"u}ller}, {Chini},
  {Schulz}, {Coulson}, {Hippelein}, {Wilke}, {Albrecht}, \&
  {Lemke}}]{Klaas2001}
{Klaas}, U., {Haas}, M., {M{\"u}ller}, S.~A.~H., {et~al.} 2001, \aap, 379, 823

\bibitem[{{Lidz} {et~al.}(2011){Lidz}, {Furlanetto}, {Oh}, {Aguirre}, {Chang},
  {Dor{\'e}}, \& {Pritchard}}]{Lidz2011}
{Lidz}, A., {Furlanetto}, S.~R., {Oh}, S.~P., {et~al.} 2011, \apj, 741, 70

\bibitem[{{Magdis} {et~al.}(2012){Magdis}, {Daddi}, {B{\'e}thermin}, {Sargent},
  {Elbaz}, {Pannella}, {Dickinson}, {Dannerbauer}, {da Cunha}, {Walter},
  {Rigopoulou}, {Charmandaris}, {Hwang}, \& {Kartaltepe}}]{Magdis2012}
{Magdis}, G.~E., {Daddi}, E., {B{\'e}thermin}, M., {et~al.} 2012, \apj, 760, 6

\bibitem[{{Mu{\~n}oz} \& {Furlanetto}(2013)}]{Munoz2013}
{Mu{\~n}oz}, J.~A., \& {Furlanetto}, S.~R. 2013, arXiv:1301.0619 

\bibitem[{{Obreschkow} {et~al.}(2009){Obreschkow}, {Heywood}, {Kl{\"o}ckner},
  \& {Rawlings}}]{Obreschkow2009}
{Obreschkow}, D., {Heywood}, I., {Kl{\"o}ckner}, H.-R., \& {Rawlings}, S. 2009,
  \apj, 702, 1321

\bibitem[{{Papadopoulos} {et~al.}(2000){Papadopoulos}, {R{\"o}ttgering}, {van
  der Werf}, {Guilloteau}, {Omont}, {van Breugel}, \&
  {Tilanus}}]{Papadopoulos2000}
{Papadopoulos}, P.~P., {R{\"o}ttgering}, H.~J.~A., {van der Werf}, P.~P.,
  {et~al.} 2000, \apj, 528, 626

\bibitem[{{Righi} {et~al.}(2008){Righi}, {Hern{\'a}ndez-Monteagudo}, \&
  {Sunyaev}}]{Righi2008}
{Righi}, M., {Hern{\'a}ndez-Monteagudo}, C., \& {Sunyaev}, R.~A. 2008, \aap,
  478, 685

\bibitem[{{Rowan-Robinson} {et~al.}(1979){Rowan-Robinson}, {Negroponte}, \&
  {Silk}}]{Rowan1979}
{Rowan-Robinson}, M., {Negroponte}, J., \& {Silk}, J. 1979, \nat, 281, 635

\bibitem[{{Scoville} \& {Solomon}(1974)}]{Scoville1974}
{Scoville}, N.~Z., \& {Solomon}, P.~M. 1974, \apjl, 187, L67

\bibitem[{{Silk} \& {Spaans}(1997)}]{Silk1997}
{Silk}, J., \& {Spaans}, M. 1997, \apjl, 488, L79

\bibitem[{{Smith} {et~al.}(2012){Smith}, {Dunne}, {da Cunha}, {Rowlands},
  {Maddox}, {Gomez}, {Bonfield}, {Charlot}, {Driver}, {Popescu}, {Tuffs},
  {Dunlop}, {Jarvis}, {Seymour}, {Symeonidis}, {Baes}, {Bourne}, {Clements},
  {Cooray}, {De Zotti}, {Dye}, {Eales}, {Scott}, {Verma}, {van der Werf},
  {Andrae}, {Auld}, {Buttiglione}, {Cava}, {Dariush}, {Fritz}, {Hopwood},
  {Ibar}, {Ivison}, {Kelvin}, {Madore}, {Pohlen}, {Rigby}, {Robotham},
  {Seibert}, \& {Temi}}]{Smith2012}
{Smith}, D.~J.~B., {Dunne}, L., {da Cunha}, E., {et~al.} 2012, ArXiv e-prints

\bibitem[{{Solomon} \& {Vanden Bout}(2005)}]{Solomon2005}
{Solomon}, P.~M., \& {Vanden Bout}, P.~A. 2005, \araa, 43, 677

\bibitem[{{Spitzer}(1978)}]{Spitzer1978}
{Spitzer}, L. 1978, {Physical processes in the interstellar medium}

\bibitem[{{Tielens}(2005)}]{Tielens2005}
{Tielens}, A.~G.~G.~M. 2005, {The Physics and Chemistry of the Interstellar
  Medium}

\bibitem[{{Tielens} \& {Hollenbach}(1985)}]{Tielens1985}
{Tielens}, A.~G.~G.~M., \& {Hollenbach}, D. 1985, \apj, 291, 722

\bibitem[{{van der Tak} {et~al.}(2007){van der Tak}, {Black}, {Sch{\"o}ier},
  {Jansen}, \& {van Dishoeck}}]{vanderTak2007}
{van der Tak}, F.~F.~S., {Black}, J.~H., {Sch{\"o}ier}, F.~L., {Jansen}, D.~J.,
  \& {van Dishoeck}, E.~F. 2007, \aap, 468, 627

\bibitem[{{Wei{\ss}} {et~al.}(2005){Wei{\ss}}, {Walter}, \&
  {Scoville}}]{Weiss2005}
{Wei{\ss}}, A., {Walter}, F., \& {Scoville}, N.~Z. 2005, \aap, 438, 533

\end{thebibliography}

\end{document}